%% file: main.tex
\documentclass[%
reprint,
superscriptaddress,
amsmath,amssymb,
aps,
pra,
noeprint, 
]{revtex4-1}

\input{preamble}

\newcommand{\papertitle}{\change{Measurement of the electric dipole moment of AlCl by Stark level spectroscopy}}

\begin{document}


\title{\papertitle}

\include{authors}

\date{\today}

\begin{abstract}
We report the measurement of the electric dipole moment of aluminum monochloride (AlCl) using a cryogenic buffer-gas beam source. \change{
Using Stark shift spectroscopy, we derive values for the dipole moments in the body-fixed frame of the two lowest vibrational states for the $\Xstate$ electronic state,
$\mu_X(v''=0) = \pdmXzero$\,D and $\mu_X(v''=1) = \pdmXone$\,D, and for the $\Astate$ state, $\mu_A(v'=0) = \pdmAzero$\,D and $\mu_A(v'=1) = \pdmAone$\,D.
}
\mycancel{Our measurements provide values for the dipole moments of the two lowest vibrational states of the $\Xstate$ and the $\Astate$ electronic states.}
We also show that the {\it ab initio} calculations of the dipole moment and $T_e$ energy of AlCl are sensitive to the level of treatment of the spin-orbit orbit coupling.
\mycancel{We also show that spin-orbit coupling with an extended number of spin states is essential in the \textit{ab initio} calculation to correctly describe both the dipole moment and the $T_e$ energy of AlCl.}
We further lay out the implications of these results for astrophysical models of stellar and planetary evolution that have used a substitute value for the dipole moment of AlCl until now.
\end{abstract}

\maketitle


\xsection{Introduction}

The ability to control molecules and the interactions between them are expected to enable a vast number of applications \cite{McCarron2018,Fitch2021}, including novel quantum computing and simulation platforms \cite{DeMille2002,Yelin2006,Yu2019,Carr2009,Micheli2006,Bao2022,Holland2022}, controlled quantum chemistry \cite{Krems2008,Ospelkaus2010,Ye2018,Heazlewood2021}, and tests of fundamental physics through precision measurements \cite{Andreev2018,
Cairncross2017,
Kozyryev2017a,
Hudson2011,
Kondov2019,
ACMECollaboration2014,
Fitch2021a,
Cairncross2017,
Yu2021,
Hutzler2020,
ORourke2019,
Aggarwal2018,
Uzan2003,
DeMille2008,
Chin2009,
Kajita2009,
Beloy2010,
Jansen2014,
Dapra2016,
Norrgard2019,
Kobayashi2019,
Chupp2019,
Kozyryev2021}.
Particularly intriguing for these endeavors are polar molecules since they offer a long-range interaction due to their electric dipole moment, which leads to interaction ranges that are 3-4 orders of magnitude larger than those of the strongest magnetic atoms \cite{PerezRios2020}.
Thus, it is of importance for the development of novel applications to have an accurate knowledge of the molecular properties of interest.
However, this is often not the case, even for diatomic molecules. For instance, no experimental data is available on the dipolar interaction strength of aluminum monochloride (AlCl). 
In fact, the question of the dipole moment of AlCl was first addressed by Lide in 1965 but was considered too complex to analyze, and only a range of $1-2$\,D was assigned \cite{Lide1965}. Consequently, any subsequent study used the average value of $1.5$\,D as a best-guess substitute in their analyses.

In this work, we report the measurement and {\it ab initio} calculations of the permanent electric dipole moment (PDM) of AlCl.
AlCl is a promising candidate for cold molecule applications that require dense molecular samples \cite{Daniel2023}.
Our analysis characterizes its interaction strength, which is required to accordingly develop, e.g.~, experiments with AlCl in optical lattices in the future. Our comparison with {\it ab initio} theory determined that the inclusion of spin-orbit interaction is required to correctly describe AlCl.
\mycancel{
Specifically, additional spin states are required in the ab initio spin-orbit treatment to accurately calculate both the PDM and electronic transition energies ($T_e$) as well as give the correct dependence of the PDM on the interncuclear distance between Al and Cl.
}
Moreover, our results differ by $\approx\!13$\,\% from the previously used substitute value, which directly feeds into astrophysical models of stellar evolution and changes the assumed limits for the observability of AlCl in distance stars, as discussed in the following.



Both Al and Cl are created in stars through nucleosynthesis processes—aluminium predominantly through fusion reactions in high-mass stars, and Cl as a byproduct of processes such as the neon burning stage \cite{Andreazza2018,Chen2024,Jones2024}. In the cooler outer layers of stars, most particularly the atmospheres of asymptotic giant branch (AGB) stars, these elements can combine to form AlCl in regions where temperatures and pressures allow molecular bonding \cite{Tsuji1973,Sharp1990}. 
AGB stars are in a late stage of stellar evolution, characterized by significant mass loss and complex chemistry in their outer layers \cite{Bloecker1995,Cernicharo2015,Kane2023}. 
The detection of AlCl in these stars therefore provides clues toward stellar formation, chemical evolution, mass loss processes, and the distribution of elemental abundances \cite{Asplund2009}.
The most notable transitions to detect AlCl include the rotational transitions near 277\,GHz and vibrational transitions around 12.7\,microns in the infrared spectrum. 
Instruments like ground-based telescopes with infrared capabilities or space-based observatories have been able to detect these lines, confirming the existence of AlCl in stellar atmospheres \cite{Cernicharo1987,Decin2017}. 
Furthermore, the ratio of Al to Cl can offer insights into the star’s nucleosynthesis history and the nature of the material from which the star formed. Spectroscopic monitoring of stars via precision radial velocity surveys also provide vast datasets to establish possible time variability of such spectral features \cite{Fischer2016,Halverson2016,Rubenzahl2023}. However, a complete analysis of AlCl in stellar spectra has previously been hampered by limited data concerning collisional excitation states and the dipole moment \cite{Ford2004}.

The relevance of AlCl extends beyond the death of stars and into the formation of new planetary systems. In circumstellar disks around young stars, Al and Cl are key components of the dust and gas that coalesce into planets \cite{Ziurys2006}. The detection of AlCl in these environments provides evidence of the chemical diversity and complexity in the material that forms planets \cite{Agundez2012}. In particular, Al is a refractory element, meaning it condenses into solid particles at high temperatures, making it a key constituent of rocky planets and asteroids \cite{Anders1989}. 
Understanding the chemical forms of Al, including its molecular bonds like AlCl, can help explain how elements are distributed in the early stages of planetary formation \cite{Lee2018}.

In the atmospheres of hot exoplanets, AlCl may play a role in the high-temperature chemistry, particularly in the interaction between metals and halogens. Though Al tends to condense out at lower temperatures, hot planetary atmospheres can result in the formation of AlCl that remains in a gaseous phase. 
This can yield information for the vertical temperature gradients in exoplanet atmospheres, such as stratification in the atmosphere, where different chemical species are confined to specific temperature zones \cite{Fortney2010,Heng2015,Madhusudhan2019}. For example, WASP-43b and HD 189733b are prime candidates for the detection of AlCl, given their high temperatures and the presence of other metal species \cite{Bouchy2005,Hellier2011,Inglis2024,Teinturier2024}. The Mid-Infrared Instrument (MIRI) on the James Webb Space Telescope (JWST) can cover the wavelength ranges necessary to detect the vibrational transitions of AlCl, which are expected to occur around 10--15\,$\mu$m \cite{Argyriou2023}. 
The strength of the absorption features, particularly at IR wavelengths, are sensitive to the dipole moment due to the effects on the vibrational modes. \change{Specifically, the absorption flux is approximately proportional to the square of the dipole moment for the molecular progenitor.} Thus, an increase in the measured dipole moment of AlCl will result in a \change{significantly} larger expected intensity in the absorption peak of the stellar spectra. Depending on the brightness and opacity of the star (such as giant stars), this may produce a measurable effect in precision measurements acquired with facilities such as JWST.


\begin{widetext}
\begin{minipage}{\linewidth}
\begin{figure}[H]
\begin{tikzpicture}

\node at (0,0) {\includegraphics[width=0.95\linewidth]{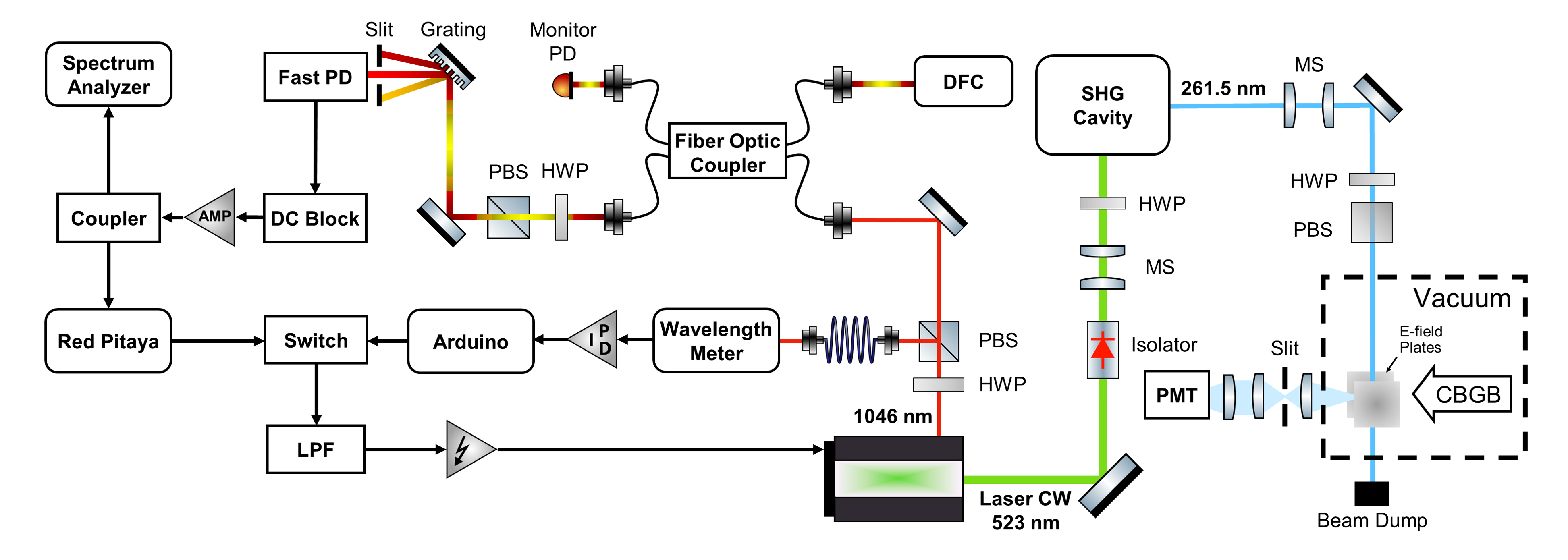}};



\end{tikzpicture}
\vspace*{-0.5cm}
\caption{
\label{fig:experimental_setup}
Experimental setup to measure the dipole moment of AlCl. A CBGB source produces a beam of AlCl that is subject to a DC electric field. The excitation laser at 261.5\,nm is frequency-stabilized using a wavelength meter, which itself is calibrated against an optical frequency comb at regular intervals.
The beat node for the comparison to the comb is produced by overlapping the lasers in a fiber coupler \company{Thorlabs, TW1064R5A2B} and sending the output on a grating \company{Thorlabs, GR25-1210}.
MS: Mode-shaping optics,
LPF: Low-pass filter,
PBS: Polarizing beam splitter,
SHG: Second-harmonic generation,
DFC: Difference frequency comb,
HWP: Half-wave plate,
PD: Photodiode,
PMT: Photomultiplier tube.
}
\end{figure}
\end{minipage}
\end{widetext}

\xsection{Methods}
A schematic of our experimental setup to measure the dipole moment of AlCl is displayed in \rfig{fig:experimental_setup}. We produce a cryogenic buffer gas beam (CBGB) \cite{Hutzler2012,Barry2011} of AlCl by ablating a powdered pressed target of Al:KCl inside a helium buffer gas cell. \change{The cell has an aperture size of $5$\,mm and is held at $\approx\,3.4$\,K by means of a two-stage pulse tube cryocooler \company{Cryomech, PT-420}.
For the ablation, we focus a pulsed Nd:YAG laser \company{Mini-Lite II, Continuum)} with an energy of $\approx\,10$\,mJ per 5\,ns pulse at a repetition rate of $2$\,Hz at $532$\,nm onto the target. The repetition rate is chosen such that the heat load on the cell is compatible with the cooling power of the fridge and the time for taking a spectrum is reasonable.
This source provides a beam of AlCl with a brightness of $\approx\,10^9$ molecules per sr per second.
}
Further details on our CBGB source, ablation technique, and vacuum apparatus are found in references \cite{Daniel2021,Lewis2021}.
At a distance of $70$\,cm downstream from the beam source, laser-induced fluorescence spectroscopy is performed on the $\Xstate \leftarrow \Astate$ transition at $261.5$\,nm with a laser beam that is aligned perpendicular to the molecular beam.
The fluorescence light is collected with a plano-convex lens with focal length $75$\,mm and imaged onto an iris for spatial filtering before being focused onto a photomultiplier tube \company{Hamamatsu, H10722-04} with two plano-convex lenses of focal length $100$\,mm.
A colored glass bandpass filter \company{Thorlabs, FGUV5M} with a transmission of 70\% at 260\,nm is added to the imaging system to minimize residual background light.

The 261.5\,nm laser light is produced in a second-harmonic generation cavity with a \change{cesium lithium borate (CsLiB6O10, CLBO)} crystal that is pumped with a $523$\,nm CW laser source \company{Vexlum, VALO SHG} and operated at a duty cycle of 5\%, which covers the transit time of molecules in the detection region, to minimize UV-induced damage on the optical components \cite{Liu2024}.
The output of the cavity is collimated and guided into the vacuum chamber to interact with the molecular beam.
For the measurements, a power of $\approx\,1$\,mW with a beam size of $\approx\,1.5$\,mm is used.
The pump laser is frequency-stabilized by referencing the pump laser to a wavelength meter \change{\company{High Finesse, WS-7}} and feeding back onto the laser piezo with an Arduino micro-controller. The wavelength meter is regularly calibrated by briefly locking the pump laser to an optical frequency comb, which is disciplined to the global positioning system.
In the calibration cycle, the lock is implemented by measuring the beat node of the pump laser and the comb, which is set to $23$\,MHz, with the \change{Fast-Fourier transform} module of a Red Pitaya microcontroller and feeding back onto the laser piezo.
We use the frequency stabilization to the wavelength meter for the experimental scans due to their large dynamic range of GHz, which is necessary for the measurement presented here.

To measure the Stark effect, the spectroscopy of the molecular beam is carried out in the presence of an electric field.
The field is produced by applying a voltage to two opposing aluminum plates of size $3 \times 3$\,cm that are held at a distance of $1$\,cm with polyether ether ketone spacers.
The molecular beam is imaged in the center between the plates, which are arranged such that the electric field is perpendicular to the beam axis. The laser beam polarization for this measurement is selected to be parallel to the electric field by means of a polarizing beam splitter.
The electric field is calibrated by measuring the polarizability of the $^2S_{1/2} - ^2P_{3/2}$ transition in potassium atoms, which are a byproduct of ablating the Al:KCl mixture target (see Supplemental Material).


\xsection{Analysis}

In the presence of an external electric field $\mathbf{E}$, the molecular energy levels are shifted and split due to the interaction of the external field with the PDM, $\bm{\mu_e}$, described by the Stark Hamiltonian \cite{Brown2003}
\be
\label{eq:stark_term}
H_s
&=& 
-\bm{\mu_e} \cdot \mathbf{E}
\quad.
\ee
\change{
The remaining Hamiltonian terms include the rovibrational and hyperfine structure terms which describe the molecular energy levels of the field-free molecule. 
We calculate the matrix elements of the full Hamiltonian of each electronic state for the basis
$\ket{\Omega, J, I_\textrm{Al}, F_1, I_\textrm{Cl}, F, M}$. Here, the used spin coupling scheme starts with the sum of the projections of the electron orbital and the electron spin momentum on the internuclear axis, $\mathbf{\Omega} = \mathbf{\Lambda} + \mathbf{\Sigma}$.
The total angular momentum is then the sum of $\mathbf{\Omega}$ and the rotational angular momentum, $\mathbf{J} = \mathbf{\Omega} + \mathbf{R}$.
The hyperfine structure of both nuclei are taken into account by coupling the total angular momentum with the nuclear spin of the aluminum atom $\mathbf{F_1} = \mathbf{J} +  \mathbf{I_\textrm{Al}}$, which in turn is coupled to the nuclear spin of the chlorine atom,  $\mathbf{F} = \mathbf{F_1} + \mathbf{I_\textrm{Cl}}$.
The two nuclear spins are $I_\textrm{Al} = 5/2$ and $I_\textrm{Cl} = 3/2$.

Explicit expressions for the matrix elements and the full algebraic treatment can be found in the Supplemental Material.
For the determination of the electric dipole moment, we then diagonalize the Hamiltonians of the two electronic state manifolds separately and extract the transition energies. \mycancel{(see Supplemental Materials)}
}

The corresponding transition frequencies are then compared with the measured spectral line frequencies at different electric field amplitudes by means of a least-square fit of the data and the model.
In particular, we calculate the \change{squared} distance of each experimental frequency, $\nu_k^\textrm{exp}$, to the nearest theoretical transition frequency, $\nu^\textrm{theo}$, and sum over these distances to calculate the fit residuals,
\begin{eqnarray}
\color{changecolor}
\label{eq:fit_residuals}
\Delta_r 
= 
\sum_k \left(\nu_k^\textrm{exp} - \nu^\textrm{theo}(\mu_X, \mu_A)\right)^2
\quad.
\end{eqnarray}

\begin{figure}[H]
\begin{center}
\begin{tikzpicture}
\node at (0,0) {\includegraphics[width=0.95\linewidth]{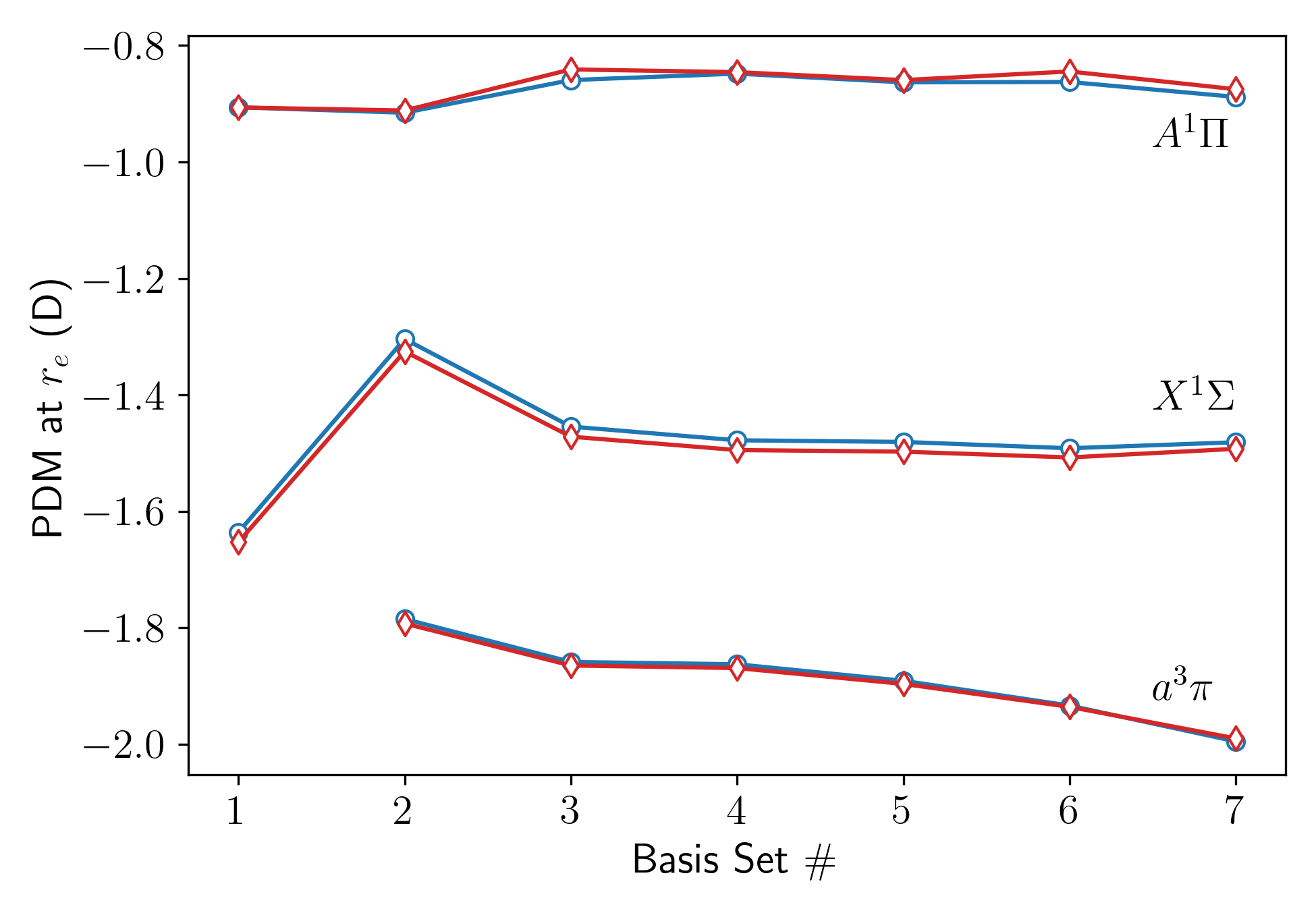}};
\end{tikzpicture}
\end{center}
\vspace*{-0.5cm}
\caption{
\label{fig:ab_initio}
\textit{Ab initio} calculation of the permanent dipole moment for the three AlCl electronic states $\Xstate$, $\Astate$, and $\apistate$ computed using seven different spin-orbit treatments and two basis sets (AV5Z blue and AVQZ red). The different treatments range from no spin-orbit (\#1), including spin-orbit (\#2), to extended spin-orbit (\#3-7).
}
\end{figure}

The fit has two free parameters, the values of the PDM $\mu_X$ for the $\Xstate$ and $\mu_A$ for the $\Astate$ state, which are determined by minimizing $\Delta_r$.
The molecular constants are taken from the literature (see \rtab{tab:alcl_molecular_constants} in the Supplemental Material). 
The measured line frequencies are determined by fitting Gaussian beam profiles to the spectral data and extracting their center frequency.
We use this method to measure the PDMs of two vibrational \change{in} both electronic states in AlCl.

In a separate effort, we have carried out complementary {\it ab initio} calculations of AlCl to determine the PDMs.
\rfig{fig:ab_initio} plots the \textit{ab initio} computed PDM for each of the three electronic states at the equilibrium internuclear distance ($R=r_e$) for several spin-orbit treatments (labeled \#1 to 7, see Supplementary Material for their detailed definitions).
Of particular interest is treatment \#1 which ignores all spin-orbit interactions, treatment \#2 which
includes spin-orbit interactions using a minimal number of spin states, and treatment \#7 which expands upon treatment \#2 by including several more spin states.
\change{
We note that while our {\it ab initio} results are in accordance with previous calculations for the different levels of spin-orbit treatment \cite{Wan2016,Yousefi2018,Bala2024,Woon2009}, the extended spin-orbit treatment is necessary to reproduce both the dipole moment and the electronic energy of the $\Astate$ (see Supplemental Material).
}
\mycancel{We note that the}
\change{The} negative PDM values are due to the chosen geometry convention that places the Cl
on the positive $z$-axis and Al at the origin. The electronegativity of Cl is larger than Al so
that the dipole points from Cl to Al (i.e., in the negative $z$ direction).


\begin{widetext}
\begin{minipage}{\linewidth}
\begin{figure}[H]
\begin{center}
\includegraphics[width=\textwidth]{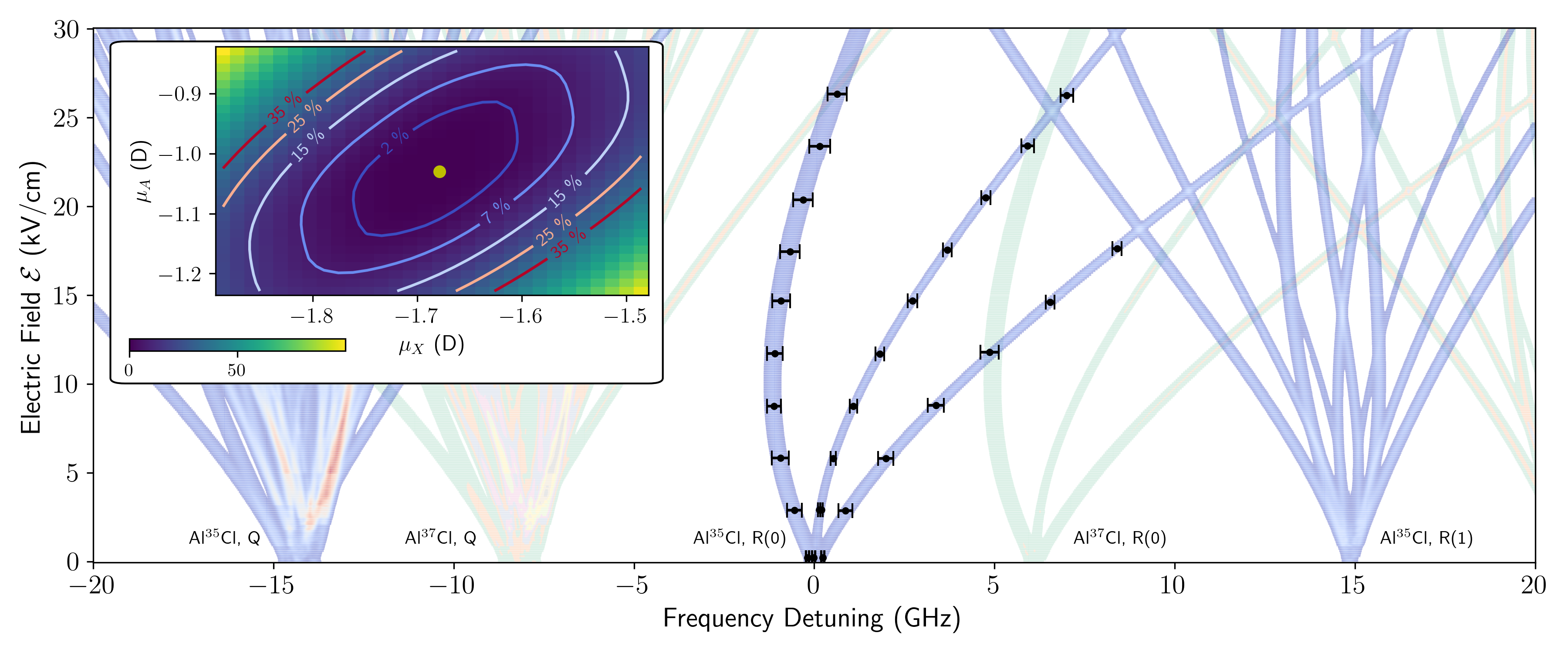}
\end{center}
\vspace*{-0.75cm}
\caption{
\label{fig:v00_results}
Stark splitting of the $\Xstate (v'' = 0) \leftarrow \Astate (v' = 0)$ transitions as a function of the laser detuning and the electric field.
The measured R(0) transition frequencies (black points) are overlayed with the theoretical predictions (blue (Al$^{35}$Cl) and green (Al$^{37}$Cl) areas) of the Stark splitting, \req{eq:stark_term}, for the optimal values in \rtab{tab:alcl_dipole_moments}.
The error bars correspond to the width of the Gaussian fits to the data.
The zero-offset detuning is set to the field-free R(0) transition frequency.
\textbf{Inset:}
\change{Normalized} residuals plot of the theory-data comparison (\req{eq:fit_residuals}) as a function of the two dipole moments, $\mu_X$ and $\mu_A$. The yellow dot marks the center-of-gravity of the \change{2}\% contour and optimal value for the PDMs.
}
\end{figure}
\end{minipage}
\end{widetext}

\xsection{Results}

\rfig{fig:v00_results} shows the comparison of the model and the measured transition frequencies as a function of the applied electric field.
To provide the full picture of the Stark splitting in AlCl, we plot the spectrum of the two main isotopologues, Al$^{35}$Cl and Al$^{37}$Cl, for the Q, R(0) and R(1) transitions.
Since both the Stark split Q and the R(1) transitions are either too dense in frequency or provide weaker signals due to lower line strengths for an unambiguous line assignment, we chose to measure the R(0) transition of the more abundant isotope, which is more isolated in frequency space, for this work.
\change{Each data point is the result of 25 averaged ablation shots and each spectral feature took approximately 12\,min to acquire.
In total, the entire data set for this work was taken over the course of several days, with an effective runtime of 9\,hours.
}

\change{
The R(0) manifold splits into three branches when subject to an external electric field since the linear component of the Stark energy in the $\Astate$ state lifts the magnetic state degeneracy.}
The Stark shift of the \change{R(0)} manifold is dominated by the value of $\mu_X$, which determines the size of the frequency shift of the rotational ground state, $J'' = 0$, in $\Xstate$. On the other hand, the size of the splitting of the manifold is governed by the value of $\mu_A$, which determines the splitting of the rotational state, $J' = 1$, in $\Astate$.
\change{
The difference in scaling of the Stark energy shift and the splitting of the $\Xstate$ and $\Astate$ states allows for determining their individual dipole moment separately by measuring the energy difference of the Stark states.
At the same time, the shift and splitting of the states are not completely independent of each other, which leads to a correlated error for the PDMs of the two electronic states.
}
\mycancel{
These two effects are not completely independent of each other, which leads to a correlated error for the PDMs of the two electronic states.
}

\begin{figure}
\begin{center}
\includegraphics[width=\linewidth]{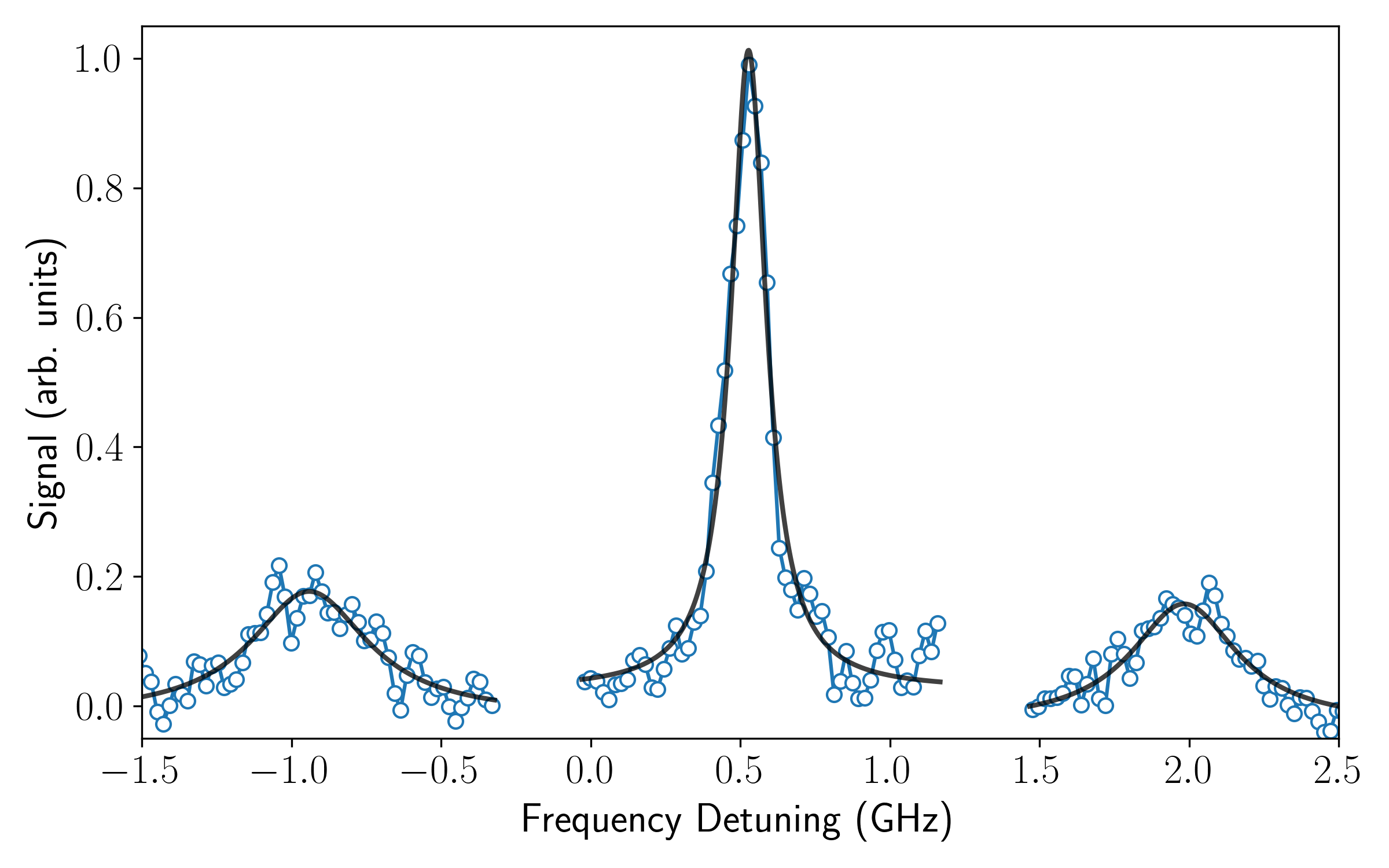}
\end{center}
\vspace*{-0.75cm}
\caption{
\label{fig:v00_line_plot}
Stark splitting at $\mathcal{E} = 5.7$\,kV/cm of Al$^{35}$Cl R(0) transition.
A moving average of two points is applied to the data and the zero offset detuning is set to the field-free transition. The black lines are Gaussian fits to extract the three peak center freqencies and the blue circles are the data.
}
\end{figure}

With the polarization parallel to the electric field, the center branch shows the strongest intensity, as shown in the example in \rfig{fig:v00_line_plot} for a field of $5.7$\,kV/cm. The linewidths of the measured peaks (FWHM: $\approx 150$\,MHz (center branch) and $\approx 300-400$\,MHz (side branches)) correspond to the theoretical expectations of the state splittings (see Supplemental Material, \rfig{fig:eigenenergies}).
We note that these broad features are the dominating source of uncertainty in fitting the transition frequencies of the data.

Under the influence of an external field, the two hyperfine spins in AlCl lead to a large number of individual transitions that \change{are} impractical to plot. Instead, for the comparison of the theory and data, we show the model as a colormap in \rfig{fig:v00_results} by summing over individual Gaussians centered at the predicted frequencies and cutting the colormap off at an intensity of 50\% for illustration purposes.
\change{
Aside from the broadening of the spectral features, for the analysis of the dipole moment, the underlying hyperfine structure can be neglected. The absence of the substructure is due to the fact that at the applied electric fields the spins decouple and the three branches of the R(0) spectrum are dominated by the Stark splitting due to the rotational states.
However, for completeness, and to provide a comprehensive description that is also valid at low electric fields for possible future experiments, we have provided the analysis that includes the hyperfine structure in the Supplemental Material.
}

\change{
The inset in \rfig{fig:v00_results} shows the residuals of the data to theory comparison of the transition frequencies. The colormap and contour lines plot the residual differences with respect to the optimal value, $\Delta_r - \Delta_r(\mu_\textrm{opt})$ and are normalized to the asymptotic value of the residuals.
We use the center of gravity of the $2$\% contour line of the residual as an estimate for the optimal value for the PDM.
}
We also carried out the same analysis for the first vibrational state $(v''=v'=1)$, see Supplemental Material.

A summary of our results of the four PDMs of AlCl and previous literature values are given in \rtab{tab:alcl_dipole_moments}.
The \textit{ab initio} PDMs in this work (bold) are consistent with previously reported PDMs.
The differences are due to the choice of the electronic basis set (the larger AV5Z in this work)
and more importantly the level of spin-orbit treatment (the extended spin basis \# 7 in this work).
A notable trend in the \textit{ab initio} PDMs for the $\Xstate$ and $\apistate$ states is the increasing magnitude
with increased vibrational excitation.  This trend is also experimentally observed for the $\Xstate$ state.
In contrast, for the $\Astate$ state the magnitudes of both the experimental and {\it ab initio}
PDMs remain about the same for $v=0$ and $v=1$.
This behavior is due to the different R dependence of the electronic dipole moments for these states
(see Supplementary Material for details).
Also of interest is the change in the PDMs with the level of spin-orbit treatment.
For the $\Xstate$ state, the $v=0$ and $v=1$ PDMs decrease significantly in magnitude when spin-orbit interactions
are included using a minimal spin basis.
When the spin basis is extended, the PDM magnitude increases becoming in better agreement the with experiment.
Similar	behavior is seen in the	$a\,^3\Pi$ state.
In contrast, for the $\Astate$ state the magnitude of the PDMs remain about the	same as	the spin-orbit	treatment is extended.

\begin{table}[ht!]
    \centering
     \caption{\label{tab:alcl_dipole_moments}
     Electric dipole moment measurement and {\it ab initio} calculation of Al$^{35}$Cl in units of Debye.
     The results from this work are listed in bold. All other listed values are previous calculations. The \textit{ab initio} results are listed without ($^\noSO$), with minimal ($^\minSO$) and with extended ($^\extSO$) spin-orbit coupling.
     }
     \begin{ruledtabular}
     \begin{tabular}{clllr}
        State & (v=0) & (v=1) & (at $r_e$)\\
        \hline
        $\Xstate$ 
        & {\bf \pdmXzero(\pdmXzeroerror)} 
        & {\bf \pdmXone(\pdmXoneerror)} & - &
        \textbf{exp}
        \\
 
         & {\bf -1.525}
         & {\bf -1.631} 
         & {\bf -1.493}
         & \textbf{\textit{ab initio}$^\extSO$}\\ 
         &  {\bf -1.319} 
         & {\bf -1.405} 
         & {\bf -1.325}
         & \textbf{\textit{ab initio}$^\minSO$}\\ 
         &  {\bf -1.648} 
         & {\bf -1.723} 
         & {\bf -1.636}
         & \textbf{\textit{ab initio}$^\noSO$}\\ 
         & - & - &-1.309 & \cite{Wan2016}$^\minSO$\\
         & -1.630 & -1.702 &-1.594 & \cite{Yousefi2018}$^\noSO$\\
         & - & - & -1.604 & \cite{Bala2024}$^\noSO$\\
         & - & - & -1.693 & \cite{Woon2009}$^\noSO$\\
         & - & - &  1--2 &  \cite{Lide1965}\\
         \\\hline
        $\Astate$ 
        & {\bf \pdmAzero(\pdmAzeroerror)} 
        & {\bf \pdmAone(\pdmAoneerror)} & - &
        \textbf{exp}
        \\
         &  {\bf -0.891} 
         & {\bf -0.890} & {\bf -0.888} 
         & \textbf{\textit{ab initio}$^\extSO$}\\ 
         &  {\bf -0.905} 
         & {\bf -0.896} & {\bf -0.916} 
         & \textbf{\textit{ab initio}$^\minSO$}\\      
         &  {\bf -0.895} 
         & {\bf -0.868} & {\bf -0.906} 
         & \textbf{\textit{ab initio}$^\noSO$}\\  
         & - & - &  -0.918 &  \cite{Wan2016}$^\minSO$\\
         & - & - &  -1.084 & \cite{Bala2024}$^\noSO$\\
         \\\hline
         
        $\apistate$ & {\bf -2.014} 
        & {\bf -2.069}
        & {\bf -1.995}
        & \textbf{\textit{ab initio}$^\extSO$}\\ 
        &{\bf -1.802} 
        & {\bf -1.845}
        & {\bf -1.785}
        & \textbf{\textit{ab initio}$^\minSO$}\\  
        & - & - & -1.760& \cite{Wan2016}$^\minSO$\\
        & - & - & -1.669 & \cite{Bala2024}$^\noSO$\\
        
      \end{tabular}
      \end{ruledtabular}
\end{table}

\xsection{Conclusion}
\label{sec:conclusion}

Our measurements conclude the outstanding question of the value of the electric dipole moment of AlCl.
\mycancel{
Our results are in excellent agreement with our ab initio calculations and with previous theoretical literature values, which are available for the vibrational ground state.}
\change{
From \rtab{tab:alcl_dipole_moments} we see that the dipole moment calculated without spin-orbit coupling shows best agreement for the PDM. However, this level of treatment gives a poor $T_e$ (see Supplementary Material, \rtab{tab:supp:pdm_diff_SO} and \rfig{fig:supp:Te_abinitio}). In contrast, including spin-orbit coupling using a minimal spin basis gives the poorest agreement for the PDM but a much better $T_e$. By extending the spin-orbit treatment, we obtain good agreement with the experimental values for {\it both} the PDM and $T_e$ simultaneously. Furthermore, the dependence of the PDM on the internuclear distance of AlCl is improved when spin-orbit coupling is included (see \rfig{fig:supp:figS2} in the Supplementary Material).
}

\mycancel{We found that including the spin-orbit coupling using an extended number of spin states is crucial for an accurate ab initio description of AlCl.
Both the PDM and $T_e$ values become in better agreement with experiment when this level of theory is used. 
The \textit{ab initio} analysis presented in this work also explains the origin for the wide range of PDM values reported in the literature.}

This result has also confirmed that AlCl is well suited for applications that require polar molecules and allows for the strategic development of such experiments with AlCl.
To the best of our knowledge, the dipole moments of the bromide and the iodide, AlBr and AlI, have only been studied theoretically or in solution \cite{Gao2017,Yang2016e,Tourky1957}.
The lightest aluminum halide, AlF, which offers excellent properties for cold molecule applications, has been studied extensively and a dipole moment of $1.515$\,D was measured \cite{Truppe2019}.

Finally, the more than 10\% increase in the value of the electric dipole moment in comparison to the previously used literature value has consequences for the expected absorption features of AlCl that may be observed for stars. \change{This translates to an increase in the expected absorption intensity, that is proportional to the square of the electric dipole moment, and is especially important at IR wavelengths where much of the current astrophysical observations occur.}
For example, the precision performance of JWST has exceeded expectations \cite{gardner2023,rigby2023}, with commissioning observations using the Near-Infrared Spectrograph (NIRSpec) instrument, which covers a wavelength range of 0.6--5.0\,$\mu$m, providing single measurement precision of 50--60\,ppm \cite{espinoza2023}. 
Early release JWST data for the hot giant planet WASP-39b resulted in the detection of the CO$_2$ absorption feature at 4.3\,$\mu$m with 26$\sigma$ significance, further demonstrating the capabilities of the instrument \cite{ahrer2023}. The expected increase in AlCl absorption may be well suited to targeted follow-up observations of the atmospheres of giant stars using the JWST/MIRI instrumentation.

\bigskip


\begin{acknowledgments}
L.~L.~, M.~A.~Jr.~and B.~H.~acknowledge funding from the National Science Foundation under Grant No.~2145147.
M.~A.~Jr.~acknowledges funding from the University of California, Riverside.
This material is based upon work supported by the Air Force Office of Scientific Research under award number FA9550-21-1-0263. B.~K.~K.~acknowledges that part of this work was done under the auspices of the U.S.~Department of Energy under Project No.~20240256ER of the Laboratory Directed Research and Development Program at Los Alamos National Laboratory. Los Alamos National Laboratory is operated by Triad National Security, LLC, for the National Nuclear Security Administration of the U.S.~Department of Energy (Contract No.~89233218CNA000001).
\end{acknowledgments}

\bigskip

\section*{Data Availability}

The data that support the findings of this article are openly available \cite{moleculeswebsite}.

\bibliography{library,stars}


\input{supplemental_material}

\end{document}

%% file: preamble.tex
\input{preamble_packages}

\input{preamble_configurations}

\input{preamble_my_bibtex}

\input{preamble_new_commands}

\input{preamble_math}

%% file: preamble_packages.tex
\usepackage{tikz}
\usetikzlibrary{arrows.meta}
\usepackage{multirow}
\usepackage{graphicx}
\usepackage{wrapfig}

\usepackage{dcolumn} 

\usepackage{bm}

\usepackage{hyperref}

\usepackage{physics}
\usepackage{float}

\usepackage{numprint}

\usepackage{xifthen}

\usepackage{xcolor}

\usepackage{epstopdf}

\usepackage{multirow}
\usepackage[normalem]{ulem}

\usepackage{chemarrow}

\usepackage{silence}
\usepackage{soul}

%% file: preamble_configurations.tex
\colorlet{tmpcolor}{black}

\colorlet{changecolor}{tmpcolor}
\colorlet{CHANGECOLOR}{tmpcolor}

\setstcolor{changecolor}

\definecolor{rulecolor}{RGB}{0,71,171}

\definecolor{tableheadcolor}{gray}{0.92}

\graphicspath{figures}
\graphicspath{{figures/}}

\hypersetup{
    colorlinks=true,
    linkcolor=blue, 
    citecolor=blue,
    urlcolor=blue
}

%% file: preamble_my_bibtex.tex
\newcommand{\aap}{Astronomy and Astrophysics}

\newcommand{\araa}{Annual Review of Astronomy and Astrophysics}

\newcommand{\apjl}{Astrophysical Journal Letters}
\newcommand{\apjs}{Astrophysical Journal Supplement}

\newcommand{\gca}{Geochimica Cosmochimica Acta}

\newcommand{\mnras}{Monthly Notices of the Royal Astronomical Society}
\newcommand{\pasp}{Publications of the Astronomical Society of the Pacific}

\newcommand{\ssr}{Space Science Reviews}

\DeclareUnicodeCharacter{3000}{\textcolor{red}{BAD!!}}

\DeclareUnicodeCharacter{2009}{\textcolor{red}{BAD!!}}

\DeclareUnicodeCharacter{03A3}{$\Sigma$}

\DeclareUnicodeCharacter{03A0}{$\Pi$}

\DeclareUnicodeCharacter{03BC}{$\mu$}

%% file: preamble_new_commands.tex
\newcommand{\rfig}[1]{Fig.~\ref{#1}}
\newcommand{\rfigs}[1]{Figs.~\ref{#1}}

\newcommand{\req}[1]{Eq.~(\ref{#1})}

\newcommand{\rtab}[1]{Tab.~\ref{#1}}

\newcommand{\company}[1]{(\!{\it #1})}

\newcommand{\Xstate}{X^1\Sigma^+}
\newcommand{\Astate}{A^1\Pi}
\newcommand{\apistate}{a^3\Pi}

\newcommand{\XAtransition}{$\Astate \leftarrow \Xstate$}

\newcommand{\alcl}[2][]{\ifthenelse{\equal{#1}{}}{Al$^{#2}$Cl}{$^{#1}$Al$^{#2}$Cl}}

\def\noSO{\diamond}
\def\minSO{\dagger}
\def\extSO{*}

\newcommand\xsection[1]{\section{#1}}

\newcommand{\mycancel}[1]{{\protect\st{#1}}}
\renewcommand{\mycancel}[1]{}

\newcommand{\change}[1]{{\color{changecolor}#1}}

\newcommand{\remove}[1]{\ignorespaces}

\def\pdmXzero{-1.679}
\def\pdmXzeroerror{78}
\def\pdmAzero{-1.030}
\def\pdmAzeroerror{111}

\def\pdmXone{-1.761}
\def\pdmXoneerror{116}
\def\pdmAone{-1.052}
\def\pdmAoneerror{149}

%% file: preamble_math.tex
\newcommand{\be}[1]{
\begin{eqnarray}#1
}

\newcommand{\ee}{
\end{eqnarray}
}

\newcommand{\ic}{i\mkern1mu}

\renewcommand{\imath}{\ic}

\newcommand{\prefac}[2]{
\sqrt{2 #1 + 1} \sqrt{2 #2 + 1}
}

\newcommand{\matelem}[3]{
\langle
#1|#2|#3
\rangle
}

\newcommand{\reducedmat}[3]{
\langle
#1||#2||#3
\rangle
}

\newcommand{\wignerthreej}[6]{
\left(
\!
\begin{array}{ccc}
  #1 & #2 & #3\\
  #4 & #5 & #6
\end{array}
\!
\right)
}

\newcommand{\wignersixj}[6]{
\left\{
\!
\begin{array}{ccc}
  #1 & #2 & #3\\
  #4 & #5 & #6
\end{array}
\!
\right\}
}



%% file: authors.tex
\author{Li-Ren Liu}%
\affiliation{Department of Physics and Astronomy, University of California, Riverside, California 92521, USA}

\author{Miguel Aguirre Jr.}%
\affiliation{Department of Physics and Astronomy, University of California, Riverside, California 92521, USA} 

\author{Stephen R. Kane}
\affiliation{Department of Earth and Planetary Sciences, University of California, Riverside, CA 92521, USA}

\author{Brian K.~Kendrick}%
\affiliation{Theoretical Division (T-1, MS B221), Los Alamos National Laboratory, Los Alamos, New Mexico 87545, USA}

\author{Boerge Hemmerling}%
\email{boergeh@ucr.edu}
\affiliation{Department of Physics and Astronomy, University of California, Riverside, California 92521, USA}

%% file: supplemental_material.tex
\clearpage

\onecolumngrid
\section{\bf Supplemental Material to ``\papertitle''}


In the following, we describe more technical details on the experiment.

\section{\change{Evaluation of the Stark energies in terms of the PDM}}

\begin{wraptable}{r}{0.3\linewidth}
\vspace*{-0.75cm}
    \centering
     \caption{\label{tab:alcl_molecular_constants}Molecular constants used in this work \cite{Daniel2023}. All units are in MHz.}
    \begin{ruledtabular}
    \begin{tabular}{clr}
        State & & \\
        \hline
        $\Xstate$ & $eQq_{0,\textrm{Al}}$ & -29.8\\
        & $eQq_{0,\textrm{Cl}}$ & -8.6\\
         \\\hline
        $\Astate$ & $a_\textrm{Al}$ & 131.9\\
        & $a_\textrm{Cl}$ & 42.0\\
        & $eQq_{0,\textrm{Al}}$ & -7.6\\
        & $eQq_{0,\textrm{Cl}}$ & -51.0\\
        & $eQq_{2,\textrm{Al}}$ & 102.9\\
        & $eQq_{2,\textrm{Cl}}$ & 32.5\\
        & $q_\Lambda$ & -3.0
    \end{tabular}
    \end{ruledtabular}
\end{wraptable}
In good approximation, the molecular states $\Xstate$ and $\Astate$ states of AlCl can be described using Hund's case (a). We briefly summarize this description here for convenience, while more details can be found in \cite{Daniel2023,Brown2003}.
\begin{minipage}{\linewidth}
\mycancel{
The used spin coupling scheme starts with the sum of the projections of the electron orbital and the electron spin momentum on the internuclear axis, $\mathbf{\Omega} = \mathbf{\Lambda} + \mathbf{\Sigma}$.
The total angular momentum is then the sum of $\mathbf{\Omega}$ and the rotational angular momentum, $\mathbf{J} = \mathbf{\Omega} + \mathbf{R}$.
The hyperfine structure of both nuclei are taken into account by coupling the total angular momentum with the nuclear spin of the aluminum atom $\mathbf{F_1} = \mathbf{J} +  \mathbf{I_\textrm{Al}}$, which in turn is coupled to the nuclear spin of the chlorine atom,  $\mathbf{F} = \mathbf{F_1} + \mathbf{I_\textrm{Cl}}$.
The two nuclear spins are $I_\textrm{Al} = 5/2$ and $I_\textrm{Cl} = 3/2$.
}
\end{minipage}
\change{We use the basis set $\ket{\Omega, J, I_\textrm{Al}, F_1, I_\textrm{Cl}, F, M}$ to describe the hyperfine states, with the spin coupling scheme as described in the main text.}
For the purpose of this work, we then diagonalize the Hamiltonian for each electronic manifold separately. The $\Xstate$ and $\Astate$ state Hamiltonians contain as dominating terms \cite{Daniel2023}
\be
\nonumber
    H_\textrm{X} &=& H_0^\textrm{X} + H_\textrm{EQ} + H_\textrm{s}^\textrm{X}\\
\nonumber
    H_\textrm{A} &=& H_0^\textrm{A} + H_\textrm{LI} + H_\Lambda + H_\textrm{EQ} + H_\textrm{s}^\textrm{A}
\ee
, where $H_0$ represents the rovibrational energy terms. 
$H_\textrm{EQ}$ is the electric quadrupole term
\begin{eqnarray}
    \nonumber
    H_\textrm{EQ} 
    &=& \sum_{\alpha = \textrm{Al, Cl}} \frac{e Q_\alpha}{4 I_\alpha (2 I_\alpha - 1)} \left[
    \sqrt{6} q_{0,\alpha} T^2_{0}(\mathbf{I}_\alpha,\mathbf{I}_\alpha)
    + \sum_{k=\pm 1} e^{(-2 \dot{\imath} k \phi)} q_{2,\alpha}
    T^2_{2 k}(\mathbf{I}_\alpha,\mathbf{I}_\alpha)
    \right]
    \quad,
\end{eqnarray}
where $e$ is the elementary electric charge, $Q_\alpha$ is the quadrupole moment of each nucleus and $q_{0,2}$ are the different components of the electric field gradients at each nucleus.

The $\Lambda$-doubling term is described by
\begin{equation}
\nonumber
    H_\Lambda = -\sum_{k=\pm 1} e^{-2 \imath k \phi} q_\Lambda T^2_{2k}(\mathbf{J}, \mathbf{J}),    
\end{equation}
where $q_\Lambda$ is the $\Lambda$-doubling constant.
The nuclear-spin-orbital hyperfine interaction is given by
\begin{equation}
\nonumber
    H_\textrm{LI} = \sum_{\alpha = \textrm{Al, Cl}} a_\alpha T^1(\mathbf{L}) \cdot T^1(\mathbf{I}_\alpha)
    \quad,
\end{equation}
where $a_\alpha$ are the orbital hyperfine constants for each nucleus.

The Stark term is derived following \cite{Brown2003}. Assuming that the electric field $\mathbf{E}$ is along the space-fixed ($p=0$) direction with amplitude $\mathcal{E}$ and assuming that the dipole moment $\bm{\mu_e}$ is along the internuclear axis ($q=0$), $\mu_0$, we express the Stark Hamiltonian in terms of spherical tensors as
\be
\nonumber
H_s
&=& 
-T^1(\mathbf{E}) T^1(\bm{\mu_e})\\
\nonumber
&=&
-T^1_{p=0}(\mathbf{E}) 
\sum_q \mathcal{D}^{(1)}_{0q}(\omega)^* T^1_{q}(\bm{\mu_e})\\
\nonumber
&=&
-\mathcal{E}
\mathcal{D}^{(1)}_{00}(\omega)^* T^1_{0}(\bm{\mu_e})
\quad.
\ee

The matrix elements of this expression are given by
\be
\nonumber
&&
-\mathcal{E}
\cdot
\matelem
{\Omega, J, I_\textrm{Al}, F_1, I_\textrm{Cl}, F, M}
{
\mathcal{D}^{(1)}_{00}(\omega)^* T^1_{0}(\bm{\mu_e})
}
{\Omega', J', I_\textrm{Al}, F_1', I_\textrm{Cl}, F', M'}\\\nonumber
&=&
-\mathcal{E}
\mu_e
(-1)^{(F-M)}
\wignerthreej{F}{1}{F'}{-M}{0}{M'}
\cdot
\reducedmat
{\Omega, J, I_\textrm{Al}, F_1, I_\textrm{Cl}, F}
{
\mathcal{D}^{(1)}_{.0}(\omega)^*
}
{\Omega', J', I_\textrm{Al}, F_1', I_\textrm{Cl}, F'}\\\nonumber
&=&
-\mathcal{E}
\mu_e
(-1)^{(F-M)}
\wignerthreej{F}{1}{F'}{-M}{0}{M'}
(-1)^{F_1 + I_\textrm{Cl} + F' + 1}
\prefac{F}{F'}\wignersixj{F_1}{F}{I_\textrm{Cl}}{F'}{F_1'}{1}\\\nonumber
&&
\quad\quad\quad\times
\reducedmat
{\Omega, J, I_\textrm{Al}, F_1}
{
\mathcal{D}^{(1)}_{.0}(\omega)^*
}
{\Omega', J', I_\textrm{Al}, F_1'}\\\nonumber
&=&
-\mathcal{E}
\mu_e
(-1)^{(F-M)}
\wignerthreej{F}{1}{F'}{-M}{0}{M'}
(-1)^{F_1 + I_\textrm{Cl} + F' + 1}
\prefac{F}{F'}\wignersixj{F_1}{F}{I_\textrm{Cl}}{F'}{F_1'}{1}\\\nonumber
&&\quad\quad\times
(-1)^{J + I_\textrm{Al} + F_1' + 1}
\prefac{F_1}{F_1'}
\wignersixj{J}{F_1}{I_\textrm{Al}}{F_1'}{J'}{1}
\reducedmat
{\Omega, J}
{
\mathcal{D}^{(1)}_{.0}(\omega)^*
}
{\Omega', J'}
\ee
, where we used the Wigner-Eckart theorem and tensor algebra to decouple the spins. The last matrix element of the Wigner rotation matrix is given by \cite{Brown2003}
\be
\nonumber
\reducedmat
{\Omega, J}
{
\mathcal{D}^{(k)}_{.q}(\omega)^*
}
{\Omega', J'}
&=&
(-1)^{J-\Omega}\prefac{J}{J'}
\wignerthreej
{J}{k}{J'}
{-\Omega}{q}{\Omega'}
\quad.
\ee

\begin{figure}[ht]
\includegraphics[width=0.85\textwidth]{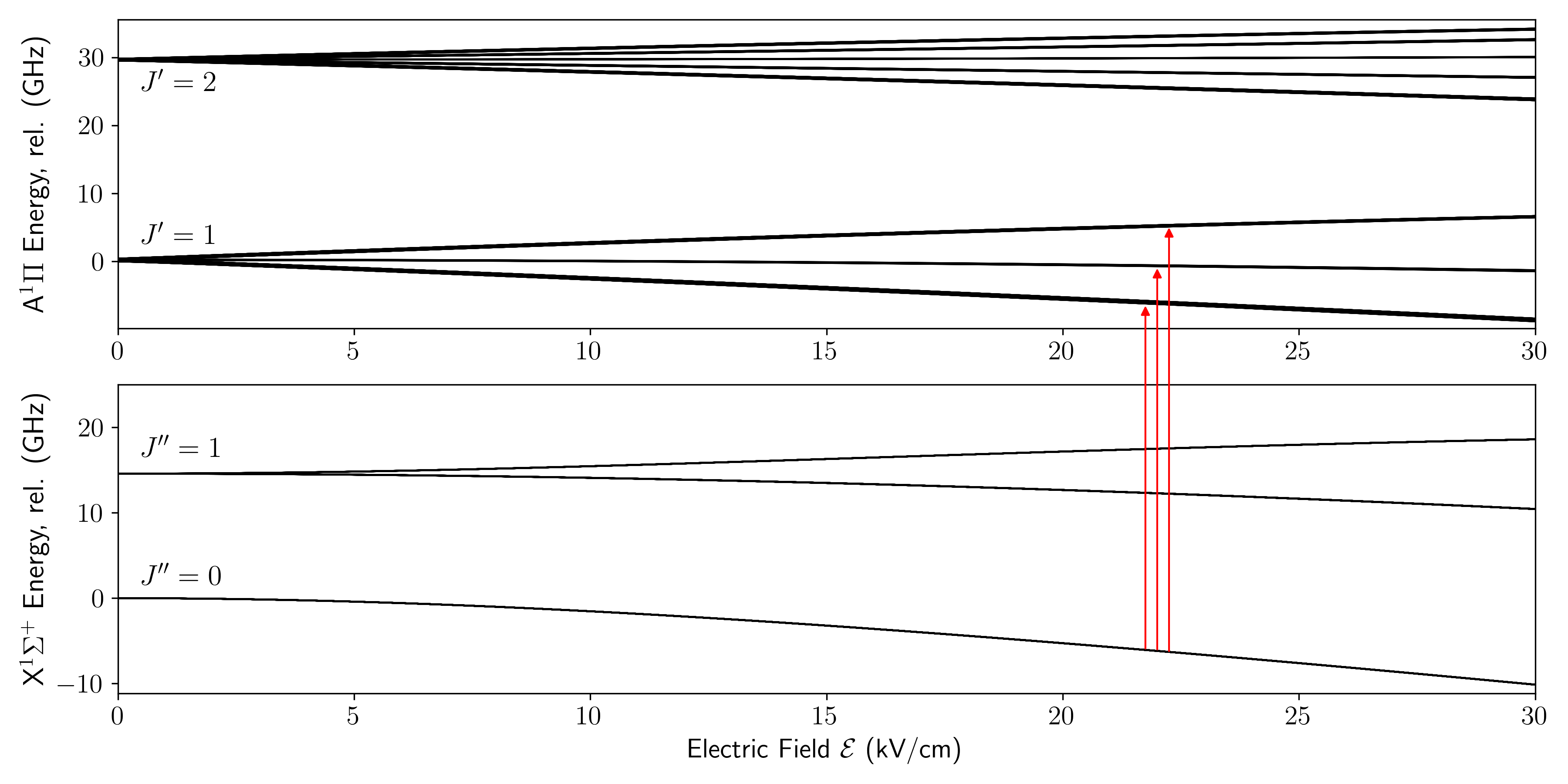}
\caption{
\label{fig:eigenenergies}
Relative energy splitting of the $\Xstate (v'' = 0)$ (lower plot) and $\Astate (v' = 0)$ (upper plot) states of AlCl as a function of the external electric field for the dipole moments $\mu_X = \pdmXzero$\,D and $\mu_A = \pdmAzero$\,D.
\change{The three red arrows, representative at $\approx 22$\,kV/cm, indicate the R(0) transitions that was used to determine the electric dipole moment of AlCl.}}
\end{figure}

The eigenenergies resulting from the diagonalization of the Hamiltonian are used to determine the transition frequencies, which are compared to the experimental data.
In \rfig{fig:eigenenergies}, the Stark shift splittings of the rotational states are shown for both electronic states for the example of the ground vibrational states $(v''=v'=0)$. 
\change{
For the $\Xstate$ state, the electric field splits up each rotational manifold into $J''+1$ number of states, whereas for the $\Astate$ state, each manifold splits into $2J'+1$ number of states in the high field regime.
This difference can readily be seen in a perturbative treatment of the Stark effect, which can be found in textbooks 
\cite{Townes1955,Brown2003}. Briefly, the $\Xstate$ state with $\Omega = 0$ has no linear, but a quadratic, contribution to the Stark energy, leading to a degeneracy of states with magnetic quantum numbers of opposite sign. On the other hand, in the $\Astate$ state, this degeneracy is lifted due to the linear Stark effect.
}
The underlying hyperfine structure is not resolved on the scale of this plot.

The three vertical red arrows illustrate an example at $\approx 22$\,kV/cm of the R(0) transitions that were measured in this work to determine the dipole moments.
The molecular constants that are used in the comparison with the data are given in \rtab{tab:alcl_molecular_constants}.

\section{Electric Field Plates and Potassium Calibration}

To produce a DC electric field, we apply a positive voltage to one of the field plates and a negative voltage to the opposite field plate, using two separate high-voltage supplies \company{ESDMC, 30kV HVM}. The actual applied voltage is monitored via a monitoring output of the power supply and recorded with each experimental sequence using our experimental control system.

While the geometry of the field plates and the applied voltage is known, we use the polarizability of potassium to calibrate the electric field that the molecules experience when traveling in between the electric field plates as a consistency check.
Potassium is a byproduct of the ablation of the Al:KCl mixture target and is thereby present in the particle beam traveling from the source to the detection region in the apparatus.
Specifically, we measure the frequency shift of the D2-line ($^2S_{1/2}, (m_J = 1/2) - ^2P_{3/2}, (m_J' = 3/2)$) and fit it to the expected frequency shift
\be
\label{eq:potassium_calibration}
\Delta \omega
=
-\frac{1}{2}
\left(
\alpha_s + \alpha_t
\frac{
3 m_J^2 - J(J+1)
}
{
J ( 2 J - 1 )
}
\right)
\mathcal{E}^2
\ee
with
$
\mathcal{E}
=
V
/
d_\textrm{eff}
$,
where $d_\textrm{eff}$ is the effective distance between the field plates and $V$ is the applied potential.

\begin{figure}
\includegraphics[width=0.4\linewidth]{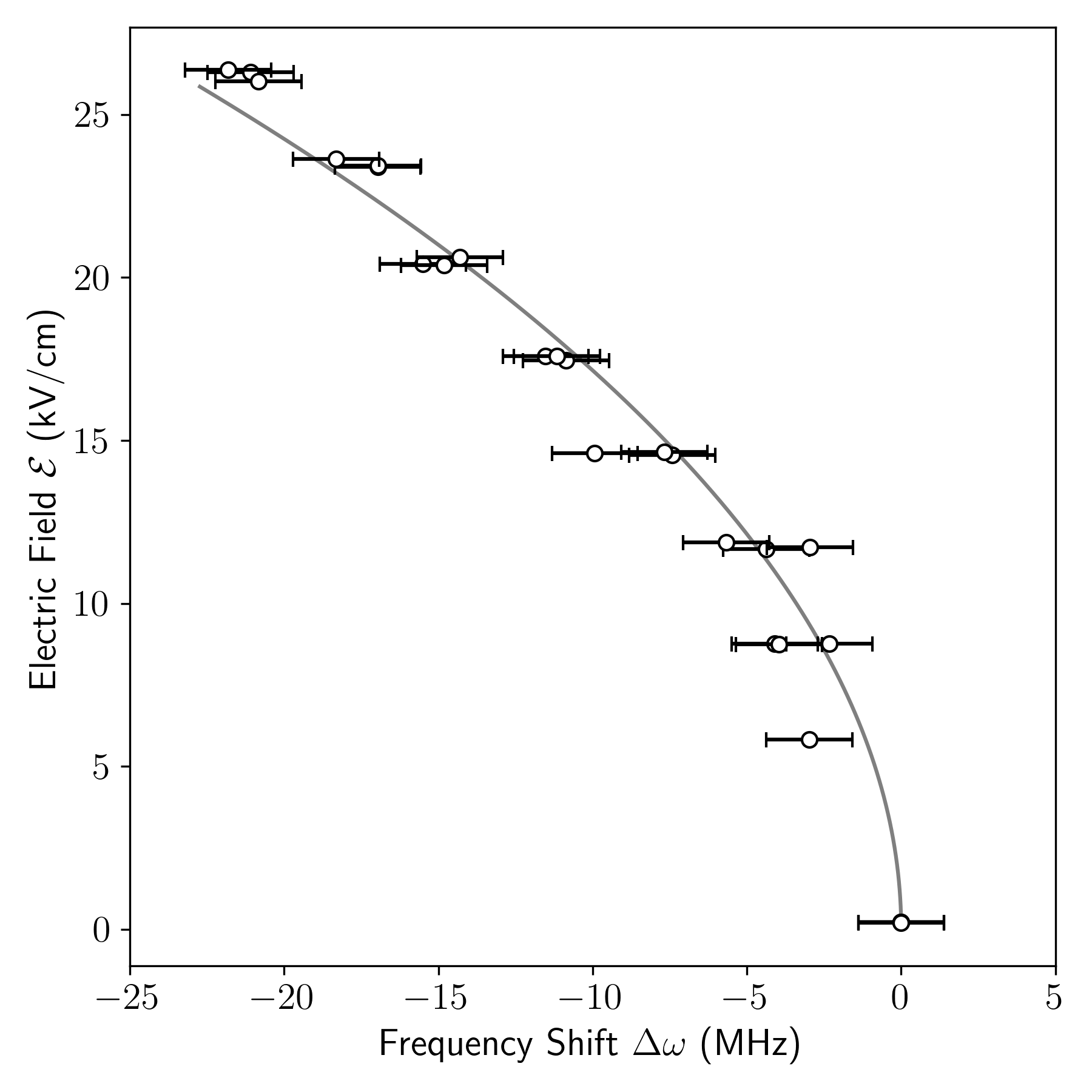}
\includegraphics[width=0.4\linewidth]{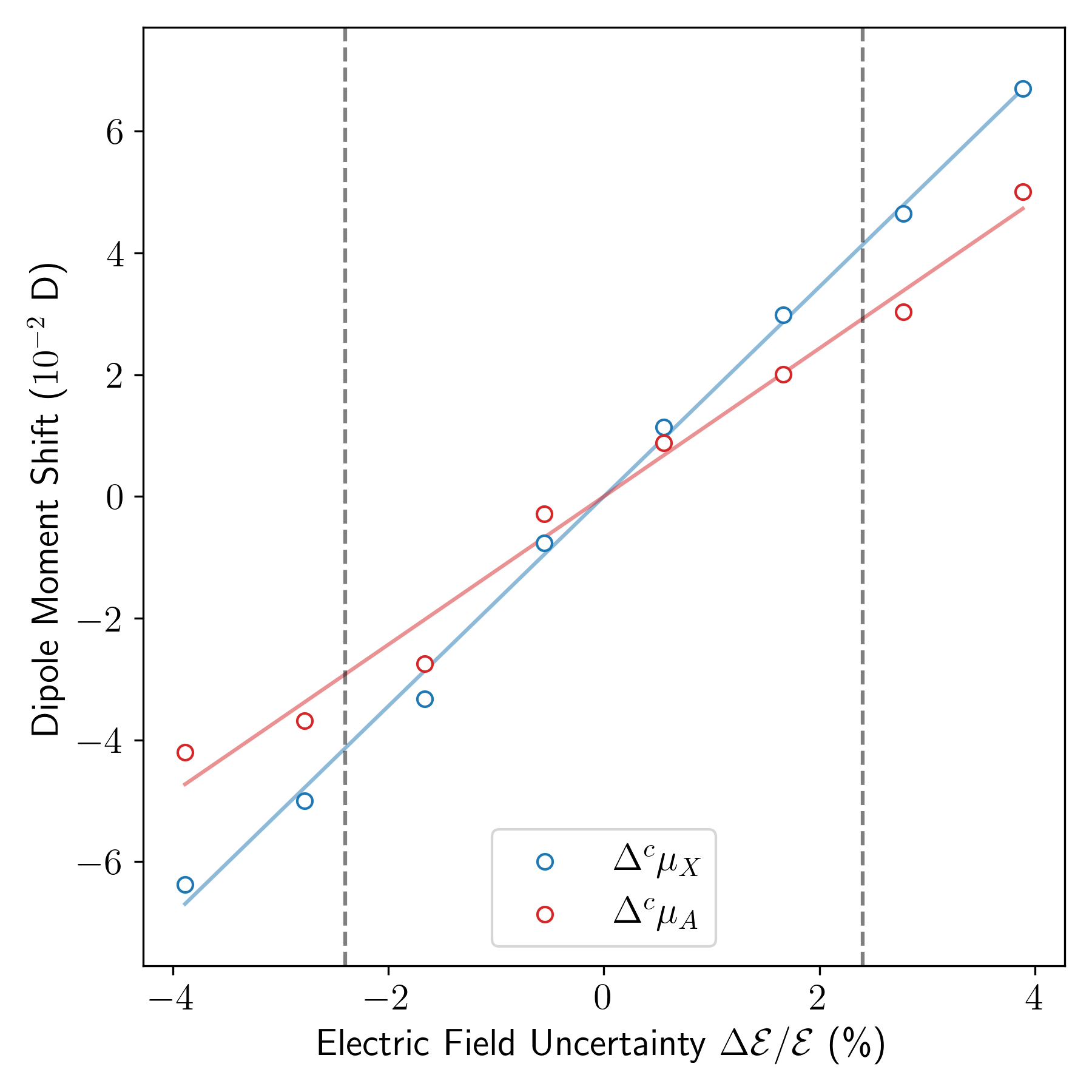}
\caption{
\label{fig:potassium_calibration}
\textbf{Left:} Frequency shift of the D2-line in Potassium as a function of the applied electric field. Blue dots are the data. Each frequency measurement has an average error of $1.4$\,MHz. The solid line is a fit to \req{eq:potassium_calibration}. The fit is anchored to the zero-field frequency.
\textbf{Right:} Shift of the electric dipole moments for both states due to an error in the value of the applied electric field. The solid lines are linear fits and the vertical dashed lines represent the uncertainty of $2.4$\%.
}
\end{figure}

Using the literature values for the scalar and tensor polarizability of potassium \cite{Kawamura2009a},
$\alpha_s(^2P_{3/2}) - \alpha_s(^2S_{1/2}) = 93.0(2.5)$\,kHz/(kV/cm$^2$) and $\alpha_t(^2P_{3/2}) = -27.6(7)$\,kHz/(kV/cm$^2$)
, a least-square fit to the data shown in \rfig{fig:potassium_calibration}(left) yields
$d_\textrm{eff} = 0.981(10)$\,cm.
This value is consistent with the geometry of the field plates, which have a nominal distance of $0.98$\,mm, and shows that we can neglect any distortion of the electric field due to the finite size of the electric field plates.

Independent of that, we estimate the effect of the error on the literature value of $\alpha_{s,t}$ on the measured dipole moments.
The uncertainty on the polarizability translates to an uncertainty of the assumed electric field as $\Delta \mathcal{E}/\mathcal{E} \approx \Delta( \alpha_s + \alpha_t)/2 (\alpha_s + \alpha_t)$.
The errors of the literature values, $\Delta \alpha_{s} = \pm 2.5$\,kHz/(kV/cm$^2$) and $\Delta 
\alpha_{t} = \pm 7$\,kHz/(kV/cm$^2$), lead to an electric field uncertainty of $\Delta \mathcal{E}/\mathcal{E} \approx 2.4\%$.

\rfig{fig:potassium_calibration}(right) shows the absolute shift of the dipole moments $\mu_X$ and $\mu_A$ resulting from our analysis of our measurements, as described in the main test, when the electric field is scaled up to $\pm 4$\%.
The resulting shift is approximately linear, with a shift of $\Delta^c \mu_X \approx 0.04$\,D and $\Delta^c \mu_A \approx 0.03$\,D at an electric field uncertainty of $\approx 2.4\%$.

We conclude that the error of the reported dipole moments, see \rtab{tab:alcl_dipole_moments}, is dominated by the finite linewidth of the R(0) transitions and errors resulting from the calibration of the electric field are at least two-fold weaker.

\section{Laser Frequency Calibration}

The frequency of the frequency-quadrupled 1046\,nm laser \company{Vexlum, VALO} driving the \XAtransition transition is continuously frequency-stabilized using a wavelength meter \company{High Finesse, WS-7}
Drifts in the wavelength meter due to environmental changes are compensated by regularly recalibrating it using the same 1046\,nm laser locked to a frequency comb \company{Toptica, DFC} for the calibration sequence.
This lock is achieved by monitoring the beat node of the frequency comb and the 1046\,nm laser with an FFT module of a Red Pitaya board, see \rfig{fig:experimental_setup}. Using a software PID module, the beat node frequency is held at a constant value by feeding back onto the piezo of the 1046\,nm laser.
During the experimental scans, the beat node is in addition recorded at each shot of the Yag on a spectrum analyzer \company{Rigol, RSA3030E}.
The frequency comb and the spectrum analyzer are referenced to the global positioning system, whose signal is distributed in the laboratory to each device by a frequency distribution amplifier \company{SRS, FS700 series}.
\change{Though the wavelength meter has a specified accuracy of only 60\,MHz, its short-term stability is by our experience much better, on the order of 5\,MHz.
We estimate that using this recalibration technique to compensate for long-term drifts limits the frequency accuracy in our measurements to $\approx \pm 10$\,MHz. This limit is a factor of $\approx 2-3$ below the natural linewidth of AlCl, which is sufficient for the purpose of this measurement.
}
\mycancel{We estimate that this recalibration technique limits the frequency accuracy to $\approx \pm 10$\,MHz, a factor of $\approx 2-3$ below the natural linewidth of AlCl, which is sufficient for the purpose of this measurement.}

\section{Spectroscopy results on the $\Xstate (v'' = 1) \leftarrow \Astate (v' = 1)$ transition}

\begin{figure}[h]
\includegraphics[width=\textwidth]{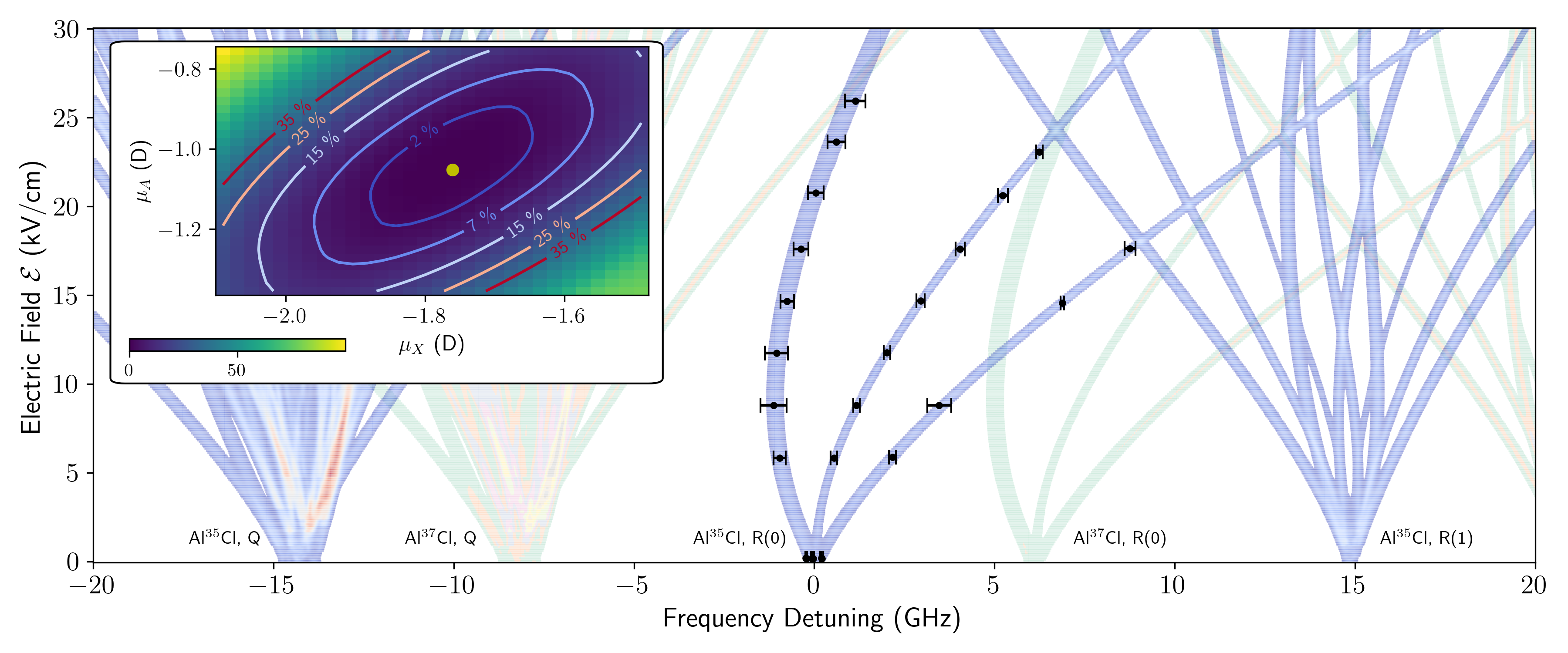}
\caption{
\label{fig:v11_results}
Stark splitting of the $\Xstate (v'' = 1) \leftarrow \Astate (v' = 1)$ transitions as a function of the laser detuning and the electric field. Only lines that could be unambiguously assigned are taken into account for the analysis. See also \rfig{fig:v00_results} for more details.
}\end{figure}

The spectroscopy measurement data on the dipole moment of the first excited vibrational states is shown in \rfig{fig:v11_results} for both electronic states, $\Xstate$ and $\Astate$. 
The analysis and the method are the same as for the lowest vibrational states and can be found in the main text and in \rtab{tab:alcl_dipole_moments}.
Again, only the spectroscopy lines in regions that show no overlap with R(1) transitions or the $^{37}$ isotope are used in the analysis for the dipole moment (black points shown in \rfig{fig:v11_results}). We note that the signal of the $v'' = 1$ transitions is a factor of $3-5$ weaker due to a lower state population in the excited vibrational state, which increases the uncertainty in the center frequency fits and leads to a slightly larger error of the PDM for these states.

\section{Discussion of the \textit{ab initio} calculations}

The \textit{ab initio} calculations for the AlCl electronic states were performed using the MOLPRO quantum chemistry code (version 2022.3) \cite{Werner2011,Werner2020,Werner2022}.
Following the approach discussed in our previous work \cite{Daniel2021}, the first steps in the MOLPRO calculations are the restricted Hartree-Fock (RHF) and complete active space self-consistent field (CASSCF) calculations using the full valence active space. The results of these calculations are then used in the final step based on the multireference configuration interaction method with the Davidson correction (MRCI+Q). A variety of electronic basis sets and
spin-orbit treatments were investigated (discussed in detail below). As part of these studies both relativistic effects using the Douglas-Kroll method and core-correlation effects using an all-electron treatment \cite{Daniel2021} were investigated. However, we found that the magnitude of the permanent dipole moment (PDM) for the $\Xstate$ state decreased slightly for the former and significantly for the later. 
Thus, including either of these effects shifts the \textit{ab initio} computed PDM further away from the measured experimental value and we therefore chose not to include them in this work.
As discussed in more detail below, we also found that using larger electronic basis sets had a relatively minor effect on the computed PDMs whereas including spin-orbit interactions and in particular extending the spin-orbit treatment to include higher spin states produced PDMs (and $T_e$) that are in overall better agreement with experiment.

\begin{figure}[h]
\includegraphics[width=0.65\textwidth]{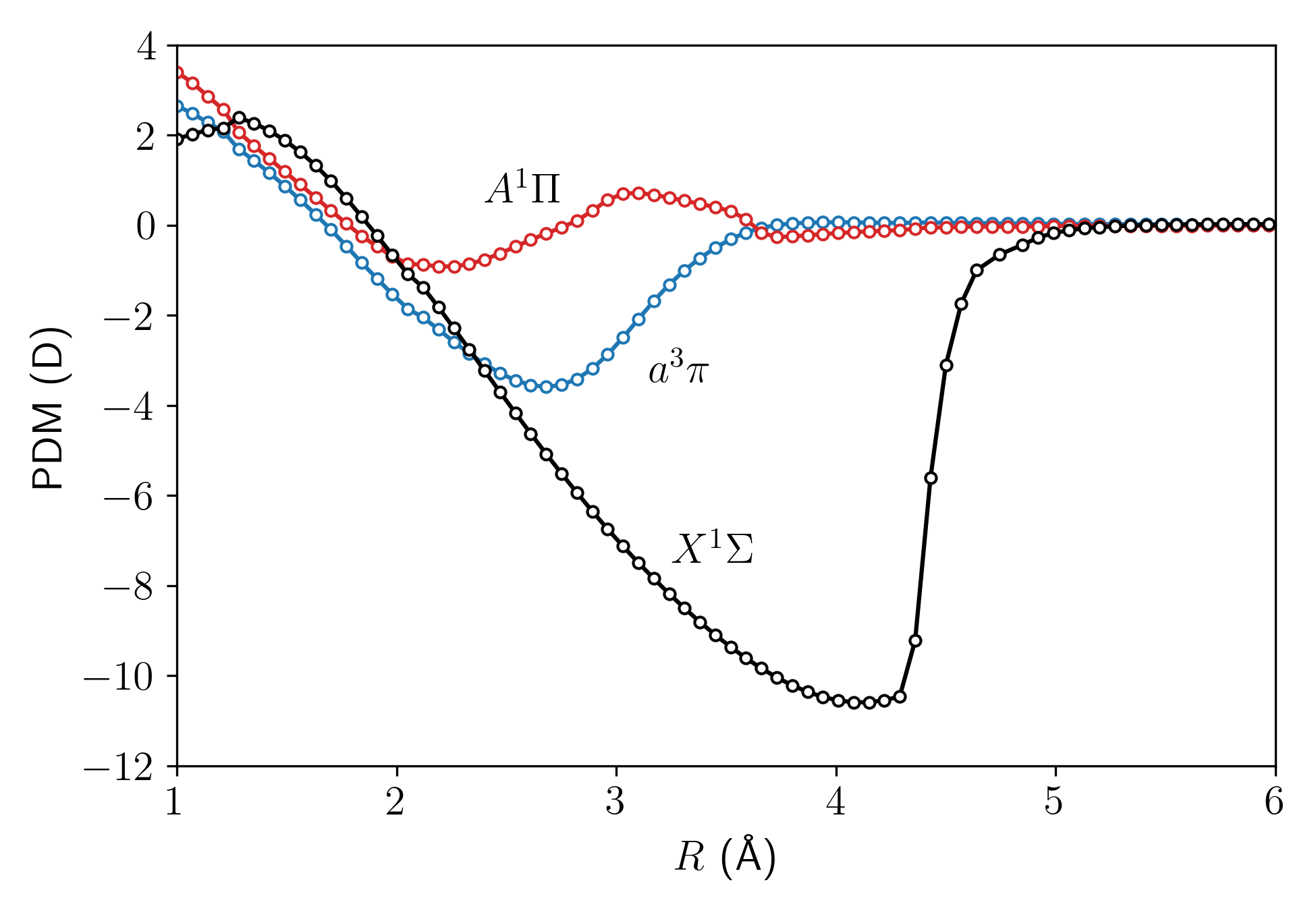}
\caption{
\label{fig:supp:pdm_ab_initio}
\textit{Ab initio} permanent dipole moments for the three AlCl electronic states $\Xstate$ (black), $\Astate$ (red), and $\apistate$ (blue) plotted as a function of internuclear distance $R$. The extended spin-orbit treatment and AV5Z basis was used and each point represents an \textit{ab initio} calculation (the points are connected by line segments to guide the eye).
}
\end{figure}

The \textit{ab initio} potential curves (electronic energies as a function of the internuclear distance $R$) for each of the three electronic states of AlCl  ($\Xstate, \Astate$ and $\apistate$) were computed for each of the electronic basis sets and spin-orbit treatments. These curves were then used in the numerical solution of the one-dimensional ro-vibrational Schr\"odinger equation (see Ref.~\cite{Daniel2021} for details). The numerically computed vibrational wavefunctions $\psi_v(R)$ are used to evaluate the dipole moment transition matrix elements $R_{vv'} = \langle \psi_v\vert\langle \phi_n\vert \mu(R) \vert\phi_n\rangle\vert\psi_v'\rangle$ where $\phi_n$ denotes the electronic wave function for a specified electronic state $n=X$, $a$, or $A$, and $\mu(R)$ is the dipole moment operator. We note that the $\langle \phi_n\vert \mu(R) \vert\phi_n\rangle$ is the PDM computed by MOLPRO as discussed above and plotted in \rfig{fig:supp:pdm_ab_initio}.
The vibrational wave functions, potential energy curves and dipole moments were fit using cubic splines and the integrals in $R_{vv'}$ over $R$ were performed using Romberg integration \cite{Press1986}.
\rtab{tab:supp:alcl_vib_pdm} lists the $\vert R_{vv'}\vert$ for each of the three electronic states and vibrational quantum numbers from $0$ to $4$.
The diagonal matrix elements $R_{00}$ and $R_{11}$ correspond to those listed in \rtab{tab:alcl_dipole_moments} of the main paper.
A notable trend in \rtab{tab:supp:alcl_vib_pdm} for the $X$ and $a$ states is the increasing $\vert R_{vv}\vert$ with increasing vibrational quantum number $v$.
This is due to the expanding width of the vibrational wavefunctions as $v$ is increased. From \rfig{fig:supp:pdm_ab_initio} we see that the PDMs for the X (black) and a (blue) states near the equilibrium values of $R=2.14$ and $2.11$\,\AA, respectively, exhibit large negative slopes. Upon vibrational excitation, the wavefunctions sample the more negative region of the PDM leading to an increase in $\vert R_{vv}\vert$.
However, for the A state we see from \rtab{tab:supp:alcl_vib_pdm} that the $\vert R_{vv}\vert$ is less sensitive and actually decreases for larger $v$.
Again from Fig. \ref{fig:supp:pdm_ab_initio} we see that the slope of the PDM for the A state (red) near the equilibrium value of $R=2.14\,$\AA\ is decreasing to zero as the PDM approaches a minimum value.
Thus, in this case it is the less negative region of the PDM that dominates the integral over R as $v$ is increased which leads to a decrease in $\vert R_{vv}\vert$.

\begin{figure}[h]
\includegraphics[width=0.65\textwidth]{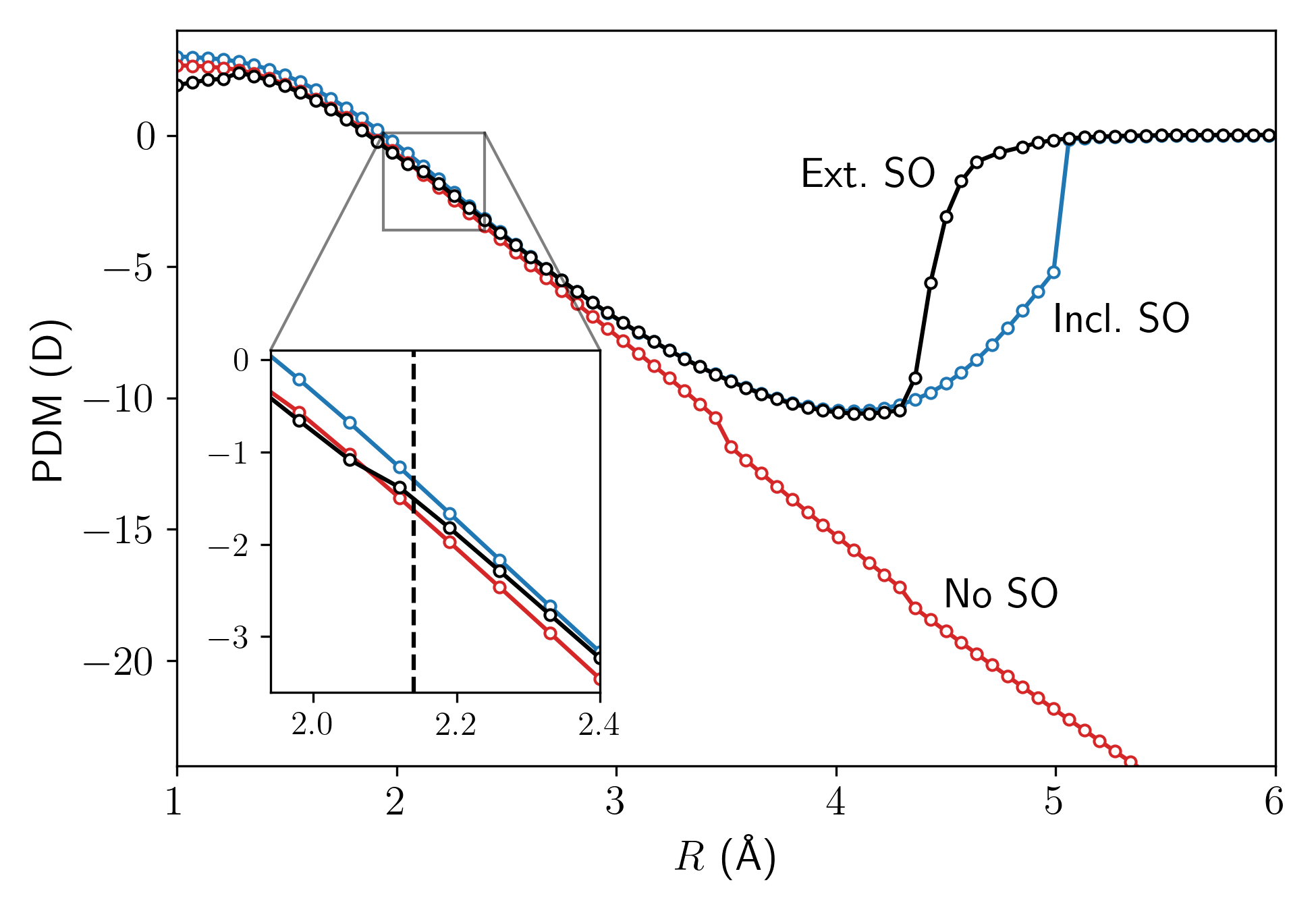}
\caption{
\label{fig:supp:figS2}
\textit{Ab initio} permanent dipole moment for $\Xstate$ plotted as a function of internuclear distance $R$. The vertical black dashed line indicates the equilibrium distance $r_e$. The AV5Z basis is used and each curve corresponds to a different spin-orbit treatment. The calculation that ignores spin orbit interactions is plotted in red whereas the blue and black data include spin orbit interactions. The black data uses an extended spin basis (i.e., \# 7 in \rfig{fig:ab_initio}).
\textbf{Inset:} Zoomed into the region near $r_e$. The range in internuclear distance is chosen to correspond to the region where the lowest lying ($v=0$) vibrational wavefunction has significant amplitude.
}
\end{figure}

The \textit{ab initio} PDMs at the equilibrium $R=r_e$ are tabulated in \rtab{tab:supp:pdm_diff_SO} for two different electronic basis sets (AV5Z and AVQZ) and several different spin-orbit treatments denoted by \# 1 to 7. These PDMs are plotted in \rfig{fig:ab_initio} of the main paper.
In what follows the different spin basis are denoted by $(i,j)$ where the $i$ denotes the irreducible representation of the $C_{2v}$ point group: $i=1$ to $4$ corresponds to $A_1$, $B_1$, $B_2$ and $A_2$, respectively \cite{Werner2022}. The $j$ denotes the spin symmetry ($2S$): $j=0$ (singlet), $1$ (doublet), $2$ (triplet), etc \cite{Werner2022}.
The seven different spin-orbit treatments are defined as: 
\begin{itemize}
\item[(\#1)] no spin-orbit interactions;
\item[(\#2)] includes spin-orbit interactions using a minimal spin basis of $(1,0)$, $(2,2)$, $(3,2)$, $(2,0)$ and $(3,0)$;
\item[(\#3)] same as \#2 above plus $(4,0)$;
\item[(\#4)] same as \#3 above plus $(4,2)$;
\item[(\#5)] same as \#4 above plus $(2,4)$;
\item[(\#6)] same as \#5 above plus $(3,4)$;
\item[(\#7)] same as \#6 above plus $(4,4)$.
\end{itemize}
For brevity, the spin-orbit treatments \#1, \#2 and \#7 are also referred to, respectively, as "without", "with minimal" and "with extended" spin-orbit coupling (as in \rtab{tab:alcl_dipole_moments} of the main paper).
From \rtab{tab:supp:pdm_diff_SO}, \rfig{fig:ab_initio} and \rtab{tab:alcl_dipole_moments} in the main paper, we see that the PDM for the X state differs significantly between the no spin-orbit (\#1) and minimal spin-orbit (\#2) treatments.
As the spin-orbit treatment is extended, the X state PDM increases significantly in magnitude from \#1 to \#2 and then more gradually from \#3 to \#7.
\rfig{fig:supp:figS2} plots the X state PDM as a function of $R$ for three different spin-orbit treatments: \#1 no spin-orbit (red), \#2 with minimal spin-orbit (blue), and \#7 with extended spin-orbit (black).
When spin-orbit interactions are ignored (red data), the X state PDM increases approximately linearly with $R$ for very large $R$.
This behavior is unphysical in that the dipole moment does not rapidly decrease to zero as the AlCl bond is broken.
Including spin-orbit interactions gives the correct behavior as seen in the blue and black data.
The inset shows a zoomed in view of the PDM near the equilibrium distance $R=r_e$ (denoted by the vertical dashed line).
The slopes of the PDM for the no spin-orbit treatment (red) and minimal spin-orbit treatment (blue) are similar.
However, the no spin-orbit PDM is more negative and therefore gives a larger $\vert R_{vv}\vert$ when integrated over $R$ (see \rtab{tab:supp:pdm_diff_SO} for $v=0$ and $1$).
The X state PDM for the extended spin-orbit treatment (black) favors the no spin orbit PDM (red) for $R<r_e$ and the minimal spin-orbit treatment (blue) for $R>r_e$.
Thus, this explains why the $\vert R_{vv}\vert$ for the extended spin-orbit treatment lies between the other two (see \rfig{fig:ab_initio} and \rtab{tab:alcl_dipole_moments}).
From \rtab{tab:supp:pdm_diff_SO}, \rfig{fig:ab_initio}, and \rtab{tab:alcl_dipole_moments} in the main paper, we see that the magnitude of the PDM for the A state increases slightly in going from the no spin-orbit (\#1) to minimal (\#2) spin-orbit treatments. Its magnitude then decreases and oscillates slightly as the spin basis is extended from \#3 to \#7.
Similarly, for the a$^3\Pi$ state we see a significant increase in the PDM magnitude between the minimal (\#2) and extended (\#3) spin-orbit treatments followed by a steady increase in magnitude from \#4 to \#7.

\begin{figure}[h]
\includegraphics[width=0.65\textwidth]{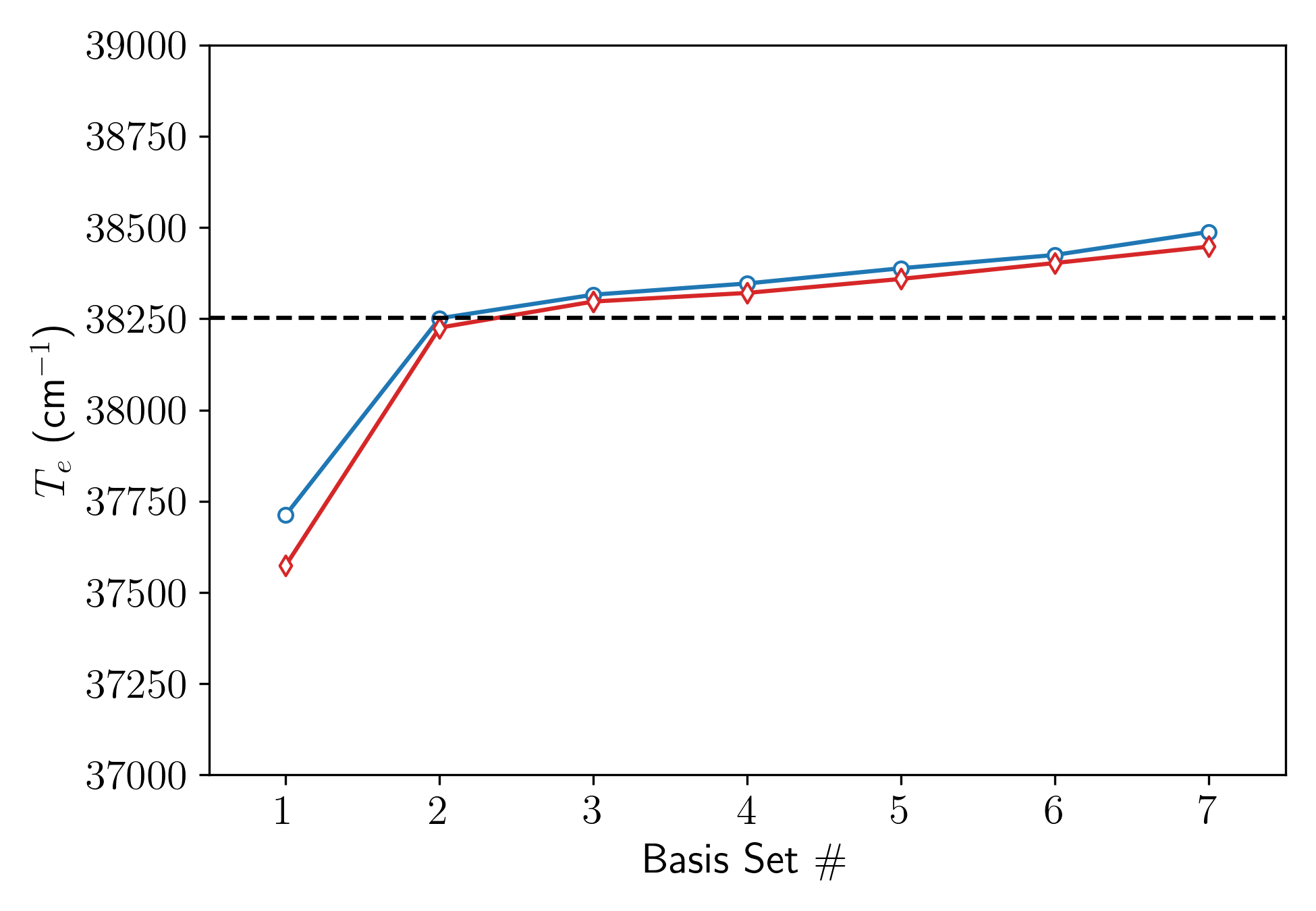}
\caption{
\label{fig:supp:Te_abinitio}
The electronic transition energy ($T_e$) for $\Astate \rightarrow \Xstate$ is plotted as a function of the spin-orbit treatment and two basis sets (AV5Z blue and AVQZ red). 
The different treatments range from no spin-orbit (\#1), including spin-orbit (\#2), to extended spin-orbit (\#3-7). The horizontal dashed black line is the experimentally measured $T_e$ \cite{Daniel2021}.
}
\end{figure}

The electronic transition energies ($T_e$) for $\Astate \rightarrow \Xstate$ listed in \rtab{tab:supp:pdm_diff_SO} are plotted in \rfig{fig:supp:Te_abinitio} as a function of the spin-orbit treatment (\#1 to \#7 defined above) and for two electronic basis sets (AV5Z in blue and AVQZ in red).
We see that including spin-orbit interactions (\#2 to \#7) gives a dramatic increase in the \textit{ab initio} computed $T_e$ (relative to the no spin-orbit treatment \#1) so that it lies much closer to the experimental value (horizontal dashed line).
The \textit{ab initio} $T_e$ increases uniformly as the spin-orbit treatment is extended (i.e., from \#2 to \#7).

For completeness, \rtab{tab:supp:pdm_fixed_re} lists the \textit{ab initio} computed PDMs and electronic energies for the X and A states near equilibrium ($R=2.1374$\,\AA) for a variety of electronic basis sets and three spin-orbit treatments (no spin-oribit (\#1), with minimal spin-orbit (\#2), and  extended spin-orbit (\#7)).
Overall, the results using different basis sets are all fairly similar.
The primary differences occur between the three different spin-orbit treatments.
In the end, we chose the AV5Z basis which seemed to give the most satisfactory results over the full range of internuclear distance R.
The larger AV6Z basis produced results very similar to AV5Z but the computational cost was significant.
The smaller AVQZ basis also produced results similar to AV5Z (e.g., compare the red and blue data in \rfigs{fig:ab_initio} and \ref{fig:supp:Te_abinitio}).

\begin{table}[h!]
    \centering
\caption{\label{tab:supp:alcl_vib_pdm}
      The vibrational transition dipole matrix elements $\vert R_{vv'}\vert$ in units of D for the $\Xstate$ state of Al$^{35}$Cl for $v$ and $v' = 0 - 4$. The dipole moment functions plotted in \rfig{fig:supp:pdm_ab_initio} were used.
}
    \begin{ruledtabular}
    \begin{tabular}{ccccccc}
    & & \multicolumn{5}{c}{$v'$}\\
    \cline{3-7}
State   & $v$   &       0    &         1       &     2     &       3       &     4\\\hline\\
  $\Xstate$  
     &    0    &    1.5255   &   0.2814    &    0.0171   &    0.0054   &   0.0098\\    
     &    1    &             &   1.6305    &    0.3992   &    0.0123   &   0.0041\\
     &    2    &             &             &    1.7201   &    0.5074   &   0.0163\\               
     &    3    &             &             &             &    1.7999   &   0.6103\\
     &    4    &             &             &             &             &   1.8803\\
\hline\\
  $\Astate$
     &    0    &    0.8907   &   0.0183    &    0.0060   &    0.0004   &   0.0050\\
     &    1    &             &   0.8894    &    0.0225   &    0.0225   &   0.0062\\
     &    2    &             &             &    0.8716   &    0.0234   &   0.0386\\
     &    3    &             &             &             &    0.8487   &   0.0176\\
     &    4    &             &             &             &             &   0.8221\\
\hline\\
  $\apistate$
     &    0    &   2.0139    &   0.1516    &    0.0047   &    0.0113   &   0.0107\\ 
     &    1    &             &   2.0691    &    0.2348   &    0.0158   &   0.0125\\
     &    2    &             &             &    2.1022   &    0.2972   &   0.0260\\
     &    3    &             &             &             &    2.1354   &   0.3492\\
     &    4	 &               &             &             &             &   2.1687\\
   \end{tabular}
    \end{ruledtabular}
\end{table}

\begin{table}[h!]
    \centering
\caption{\label{tab:supp:pdm_diff_SO}
 The \textit{ab initio} permanent dipole moment computed at $r_e$ for different spin-orbit treatments and basis sets (AV5Z and AVQZ) plotted in \rfig{fig:ab_initio}. The electronic transition energy ($T_e$) plotted in \rfig{fig:supp:Te_abinitio} is also tabulated as well as the equilibrium distances $r_e$.
}
    \begin{ruledtabular}
    \begin{tabular}{llcccccccc}   
           & & &       1     &    2     &    3     &    4     &    5     &    6     &    7\\\hline\\
\multirow{6}{*}{PDM (D)} & $\Xstate$ & AV5Z &     -1.6358  & -1.3041  & -1.4545  & -1.4781  & -1.4809  & -1.4917  & -1.4814\\
&   & AVQZ &     -1.6528  & -1.3254  & -1.4718  & -1.4948  & -1.4976  & -1.5075  & -1.4927\\
&  $\Astate$ & AV5Z &     -0.9063  & -0.9155  & -0.8594  & -0.8484  & -0.8633  & -0.8627  & -0.8884\\
&   & AVQZ &     -0.9065  & -0.9117  & -0.8412  & -0.8458  & -0.8595  & -0.8447  & -0.8752\\
& $\apistate$ & AV5Z &              & -1.7849  & -1.8587  & -1.8627  & -1.8910  & -1.9336  & -1.9952\\
& & AVQZ &              & -1.7927  & -1.8648  & -1.8692  & -1.8964  & -1.9358  & -1.9900\\           
\hline\\
\multirow{6}{*}{$r_e$ (\AA)} & $\Xstate$ &  AV5Z &      2.1404  &  2.1395  &  2.1451  &  2.1453  &  2.1450  &  2.1444  &  2.1362\\
& &  AVQZ &      2.1431  &  2.1421  &  2.1477  &  2.1478  &  2.1476  &  2.1471  &  2.1394\\
& $\Astate$ &  AV5Z &      2.1442  &  2.1343  &  2.1395  &  2.1397  &  2.1392  &  2.1382  &  2.1369\\
& &  AVQZ &      2.1477  &  2.1386  &  2.1441  &  2.1442  &  2.1436  &  2.1429  &  2.1403\\
& $\apistate$ &  AV5Z &              &  2.1038  &  2.1102  &  2.1098  &  2.1098  &  2.1090  &  2.0929\\
& &  AVQZ &              &  2.1070  &  2.1129  &  2.1131  &  2.1131  &  2.1117  &  2.0961\\
\hline\\
\multirow{4}{*}{$T_e$ (cm$^{-1}$)} & $\Astate \rightarrow \Xstate$ & AV5Z & 37712.5  & 38251.2  & 38315.9  & 38346.6  & 38388.3  & 38424.7  & 38487.9\\
& & AVQZ & 37574.2  & 38225.6  & 38297.1  & 38320.7  & 38359.1  & 38402.6  & 38447.6\\\\
& Experiment \cite{Daniel2021}  &  & 38253.22
   \end{tabular}
    \end{ruledtabular}
\end{table}

\begin{table}[h!]
    \centering
\caption{\label{tab:supp:pdm_fixed_re}
\textit{Ab initio} permanent dipole moments and energies for the $\Xstate$ and $\Astate$ states of AlCl at a fixed internuclear distance near equilibrium. The MOLPRO computed results using different basis sets and spin-orbit treatments are listed.
}
    \begin{ruledtabular}
    \begin{tabular}{lcccccc}
               \multicolumn{7}{c}{PDM (D) at $R = 2.1374$\,\AA}\\
               \hline\\
& \multicolumn{2}{c}{No Spin Orbit} 
& \multicolumn{2}{c}{Incl.~Spin Orbit} 
& \multicolumn{2}{c}{Ext.~Spin Orbit}\\
\cline{2-3} \cline{4-5} \cline{6-7}
        & X  & A   & X & A   & X  & A  \\\hline\\
VQZ     &     -1.5999   & 	   -0.8742      &        -1.2810    &     -0.9360    &      -1.4731 & 	-0.8692\\ 
AVQZ    &     -1.6142   & 	   -0.8835      &        -1.2926    &     -0.9093    &      -1.4840 & 	-0.8750\\
ACVQZ   &     -1.6140   &  	   -0.8828      &        -1.2903    &     -0.9114    &      -1.4855 &  	-0.8790\\
AWCV5Z  &     -1.6159   &      -0.8919      &        -1.2883    &     -0.9229    &      -1.4880 & 	-0.8905\\
ACV5Z   &     -1.6156   &  	   -0.8915      &        -1.2882    &     -0.9224    &   	-1.4877 &   -0.8902\\
ACV6Z   &     -1.6164   &      -0.8948      &        -1.2876    &     -0.9268    &      -1.4887 & 	-0.8946\\
AV5Z    &     -1.6156   &  	   -0.8910      &        -1.2892    &     -0.9219    &      -1.4867 & 	-0.8885\\
AV6Z    &     -1.6166   &  	   -0.8946      &        -1.2881    &     -0.9267    &      -1.4885 & 	-0.8941\\
AVQZ+d  &     -1.6142   &  	   -0.8836\\
AV5Z+d  &     -1.6155   &  	   -0.8910\\
\hline\\
 \multicolumn{7}{c}{Energies (Hartree) $R = 2.1374$\,\AA}\\
               \hline\\
& \multicolumn{2}{c}{No Spin Orbit} 
& \multicolumn{2}{c}{Incl.~Spin Orbit} 
& \multicolumn{2}{c}{Ext.~Spin Orbit}\\
\cline{2-3} \cline{4-5} \cline{6-7}
        & X  & A  & X & A  & X & A \\\hline\\
VQZ	   & -701.815618 &	-701.644112 & -701.816466 & -701.641960 & -701.816969 & -701.641500\\
AVQZ   & -701.817752 &	-701.646538 & -701.818599 & -701.644435 & -701.819126 & -701.643945\\
ACVQZ  & -701.819268 &	-701.648017 & -701.820124 & -701.645965 & -701.820642 & -701.645457\\
AWCV5Z & -701.826068 &	-701.654081 & -701.826839 & -701.652537 & -701.827322 & -701.651935\\
ACV5Z  & -701.825727 &	-701.653786 & -701.826511 & -701.652218 & -701.826999 & -701.651623\\
ACV6Z  & -701.828254 &	-701.656040 & -701.829011 & -701.654656 & -701.829487 & -701.654026\\
AV5Z   & -701.824932 &	-701.653094 & -701.825693 & -701.651407 & -701.826185 & -701.650823\\
AV6Z   & -701.827939 &	-701.655763 & -701.828684 & -701.654329 & -701.829162 & -701.653703\\
   \end{tabular}
    \end{ruledtabular}
\end{table}

